\documentclass[acmcsur]{acmsmall}

 \pdfoutput=1

\usepackage{verbatim}
\usepackage[protrusion=true,expansion=true]{microtype}
\usepackage{amsfonts,amsmath,amssymb,amstext,latexsym}
\usepackage{xspace}
\usepackage{array}
\usepackage{multirow}
\usepackage{booktabs}
\usepackage{color} 
\usepackage{ifthen} 
\usepackage{graphicx} 
\usepackage[hang,small]{caption} 
\usepackage{floatflt}
\usepackage[normalem]{ulem}
\usepackage{wrapfig}

\graphicspath{{Figures/}}

\newcommand{\itemb}{\item[$\bullet$]}

\newcommand{\ignore}[1]{}

\newcommand{\notes}[1]{~{\color{blue}{\emph{#1}}}\color{black}~}

\newcommand{\debutajout}{~\color{blue}\rule[-.1cm]{.05cm}{.4cm}~\rule[-.1cm]{.1cm}{.4cm}~\rule[-.1cm]{.2cm}{.4cm}~\rule[-.1cm]{.3cm}{.4cm}~\rule[-.1cm]{.4cm}{.4cm}}
\newcommand{\finajout}{~\rule[-.1cm]{.4cm}{.4cm}~\rule[-.1cm]{.3cm}{.4cm}~\rule[-.1cm]{.2cm}{.4cm}~\rule[-.1cm]{.1cm}{.4cm}\color{black}}

\newboolean{showcomments}
\setboolean{showcomments}{false}
\ifthenelse{\boolean{showcomments}}
{ \newcommand{\mynote}[3]{
    \fbox{\bfseries\sffamily\scriptsize#1}
    {\small$\blacktriangleright$\textsf{\emph{\color{#3}{#2}}}$\blacktriangleleft$}}}
{ \newcommand{\mynote}[3]{}}
\newcommand{\shrink}[1]{}
\definecolor{orange}{rgb}{1,0.5,0.5}

\newboolean{showchanges}
\setboolean{showchanges}{true}
\ifthenelse{\boolean{showchanges}}
{ 
	
	\newcommand{\deleted}[1]{\color{red}{\sout{#1}}\xspace\color{black}}
} {
	
	\newcommand{\deleted}[1]{}
}

\acmVolume{49}
\acmNumber{2}
\acmArticle{}
\acmYear{2016}
\acmMonth{11}

\title{Confidentiality-Preserving Publish/Subscribe: a Survey} 

\author{EMANUEL ONICA$^1$, PASCAL FELBER$^2$, HUGUES MERCIER$^2$, ETIENNE RIVI{\`E}RE$^2$\\
1: Alexandru Ioan Cuza University of Ia\c{s}i, Romania \\
2: Institut d'Informatique, Universit{\'e} de Neuch{\^a}tel, Switzerland
}

\begin{abstract}
Publish/subscribe (pub/sub) is an attractive communication paradigm for large-scale distributed applications running across multiple administrative domains.
Pub/sub allows event-based information dissemination based on constraints on the nature of the data rather than on pre-established communication channels.
It is a natural fit for deployment in untrusted environments such as public clouds linking applications across multiple sites.
However, pub/sub in untrusted environments lead to major confidentiality concerns stemming from the content-centric nature of the communications.
This survey classifies and analyzes different approaches to confidentiality preservation for pub/sub, from applications of trust and access control models to novel encryption techniques.
It provides an overview of the current challenges posed by confidentiality concerns and points to future research directions in this promising field.
\end{abstract}

\ccsdesc[500]{General and reference~Surveys and overviews}
\ccsdesc[500]{Software and its engineering~Publish-subscribe / event-based architectures}
\ccsdesc[300]{Security and privacy~Cryptography}
\ccsdesc[300]{Security and privacy~Access control}
\ccsdesc[300]{Security and privacy~Privacy-preserving protocols}
\ccsdesc[500]{Computer systems organization~Distributed architectures}

\terms{Algorithms, Security, Design}
\keywords{publish/subscribe, confidentiality}

\begin{document}

\begin{bottomstuff}
\copyright ACM, 2016. This is the author's version of the work. It is posted here by permission of ACM for your personal use. Not for redistribution. The definitive version was published in ACM Computing Surveys, Volume 49, Issue 2, (November 2016), http://doi.acm.org/10.1145/2940296
\end{bottomstuff}

\maketitle


\section{Introduction}
\label{sec:introduction}

\begin{figure}[]
\centering
\includegraphics[scale=0.5]{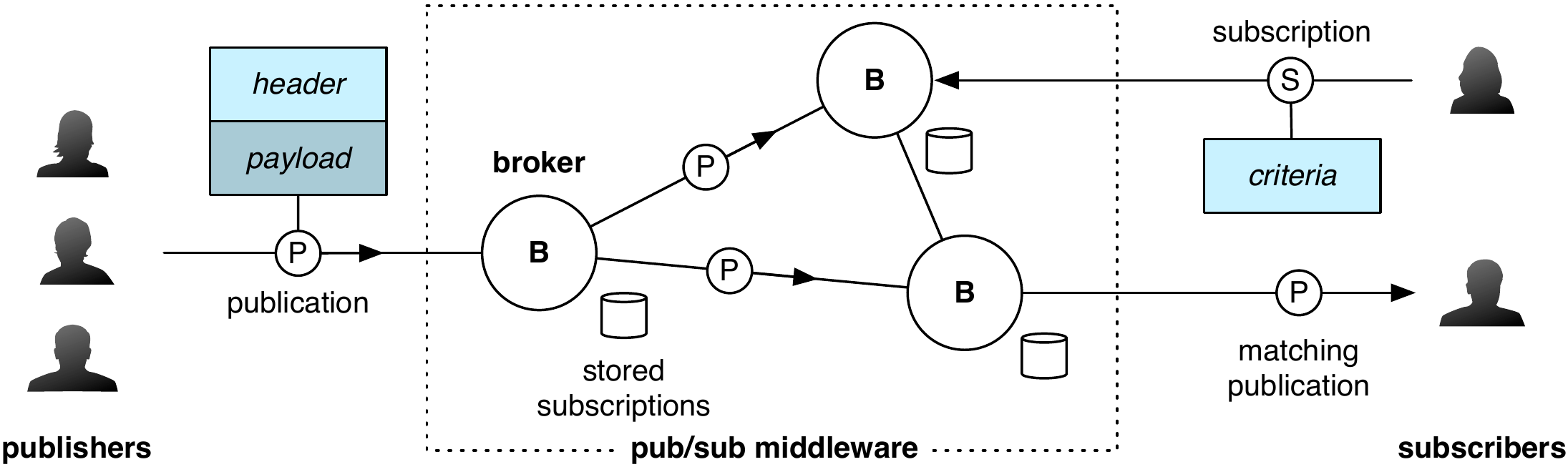}
\caption{A generic broker-based pub/sub system.}
\label{fig.pubsub}
\end{figure}

Publish/Subscribe (\emph{pub/sub}) systems~\cite{Eugster2003The-many-faces-of-pu} allow disseminating information in distributed systems from several sources (the \emph{publishers}) to different subsets of interested users (the \emph{subscribers}).
Publishers produce data in the form of \emph{publications}.
Subscribers express their interests for receiving a subset of publications by issuing \emph{subscriptions} composed of predicates, or \emph{constraints}.
Any publication \emph{matching} a given subscription's constraints is delivered to the corresponding subscriber.
The most common approach in pub/sub systems is to consider that the matching procedure is performed by a set of dedicated machines, the \emph{brokers}.
The brokers, typically organized in an overlay, store the subscriptions received from subscribers and filter incoming publications, which are forwarded to the interested subscribers. 
Communication between publishers and subscribers is decoupled in time and space.
Publishers do not need to know the identity of the interested subscribers, nor do they need to synchronize with them.
The task of determining the subset of interested subscribers and routing the publications is the responsibility of the pub/sub system itself.
A generic broker-based pub/sub system is illustrated in Figure~\ref{fig.pubsub}.

\subsection{Pub/sub paradigms}
\label{sec:intro:paradigms}

Pub/sub systems are typically classified according to the model they allow for subscriptions constraints.
The two main models are \emph{topic-based} and \emph{content-based}.

In topic-based pub/sub, subscribers declare one or several topics of interest among a list of predefined topics.
Topic-based pub/sub is a form of group communication.
A publication is tagged with a topic and propagated to all the subscribers that registered a subscription for that topic.
The main drawback of topic-based pub/sub is its low expressiveness\footnote{The capacity of a query language to express more or less complex queries (e.g., conjunctions vs. simple terms, multi-dimensionality, etc.).} caused by this use of predefined topics.
Implementations of this model can however often be optimized for high throughput and scalability, e.g., by pre-organizing subscribers for the same topic in a distributed dissemination structure~\cite{Castro2002Scribe:-A-large-scal,Patel2009Rappel:-Exploiting-i,zeromq}.
Examples of topic-based pub/sub systems include Apache Hedwig~\cite{Apache-HedWig}, Bayeux~\cite{Zhuang2001Bayeux:-an-architect}, MQTT~\cite{mqtt}, Rappel~\cite{Patel2009Rappel:-Exploiting-i}, Scribe~\cite{Castro2002Scribe:-A-large-scal}, Sprinkler~\cite{Geng2013Sprinkler---Rel}, TIB/RendezVous~\cite{Oki1993The-Information} and ZeroMQ~\cite{zeromq}.

In content-based pub/sub, the set of interested subscribers is determined at runtime based on the content of publications.
The content is typically summarized or represented by a \emph{header} that contains a collection of values over some \emph{attributes}.
Subscriptions can filter publications of interest via a set of predicates, which are constraints over these attributes.
This paradigm is strictly more expressive than topic-based pub/sub: 
a subscriber is no longer limited to a predefined topic and can combine different types of constraints on any of the attributes.
Examples of content-based pub/sub include Elvin~\cite{Segall2000Content-Based-R}, Gryphon~\cite{Strom1998Gryphon:-An-Inf,Team04Achieving-ScalabilitWD}, Hermes~\cite{Pietzuch2002Hermes:-A-Distribute}, JEDI~\cite{Cugola2001The-JEDI-Event-}, PADRES~\cite{Li2005A-Unified-Appro,Jacobsen2009The-PADRES-Publ}, Rebecca~\cite{Muhl2001Generic-Constra,Muhl2002Large-Scale-Con}, Siena~\cite{Carzaniga2001Design-and-Evaluatio}, StreamHub~\cite{Barazzutti2013StreamHub:-A-Ma,Barazzutti:2014db} and XNET~\cite{Chand2004XNet:-a-Reliable-Con}.

As topic-based pub/sub can be modeled as a special and simpler case of content-based pub/sub where the allowed constraint is a single equality over a single attribute, we will use the term \emph{constraint(s)} to denote either cases in the remainder of this section (unless otherwise specified).

The expressiveness and flexibility of pub/sub has led to broad range of applications. 
Applications include the dissemination of stock quotes~\cite{Machanavajjhala2008Scalable-ranked}, 
E-Health information systems~\cite{Ion2010An-implementati,Eze:2010:PDI:1807514.1807520}, network management systems~\cite{Martin-Flatin1999A-Survey-of-Dis,PereraA-Scalable-and-}, RSS feed monitoring~\cite{Rose2007Cobra:-Content-}, and algorithmic trading with complex event processing~\cite{PIETZUCH2004Composite-event,Adi2006Complex-Event-P}.

\subsection{Pub/sub confidentiality}
\label{sec:intro:case}

The focus of this survey is on \emph{confidentiality}, one of the main security concerns arising in multi-party pub/sub networks, and more generally in communication networks where untrusted parties must do some computation on sensitive data.
This is of increasing importance with the advent of externalized computing resources (in particular, cloud computing) and the associated possibility to offer pub/sub \emph{as a service}. 
We exemplify confidentiality challenges with two seminal applications.

Our first application is a stock quote notification system, and is widely used in the literature (e.g.,~\cite{Machanavajjhala2008Scalable-ranked}).
Publishers are stock markets issuing trading quotes.
Subscribers are investors or other financial institutions wishing to receive quotes according to their various interests, such as all quotes above a certain volume of exchange, quotes with the highest daily variation, etc.
A set of agencies provides pub/sub services by filtering quotes received from publishers according to the constraints defined by subscribers. 
In this context, publications are public data, but subscriptions may contain highly sensitive information.
Indeed, leaking the subscriptions originating from a customer could reveal information about its investment strategies and be used by competitors.
We point out that the encryption of subscriptions using standard techniques before submission to the broker is inadequate: the pub/sub system must be able to route publications based on the constraints set in these subscriptions.
If the constraints are encrypted such that they are completely opaque to the pub/sub system, routing is impossible and all filtering will have to be performed at the subscribers' side.
Ensuring confidentiality in such a system is therefore a compromise between the ability to accurately route publications and the risk of leaking information.

E-Health information systems~\cite{Ion2010An-implementati,Eze:2010:PDI:1807514.1807520} is another application that can benefit from content-based pub/sub as a communication layer.
Such systems link actors of public and private health sectors (physicians, hospitals, clinics, pharmacists).
These actors share files about patients to ensure a timely dissemination of cases, tests, etc.
A typical publisher could be an emergency unit receiving persons in critical condition. 
In this case, publications include the identity of the patients along with the content of their medical files. 
This information must be disseminated to various hospital units, possibly geographically separated and in independent administrative domains, where the patient can be moved when his condition stabilizes or where tests have to be performed or analyzed.
These healthcare units can submit subscriptions to the pub/sub system to take act of new cases, organize and schedule the patient admission and treatment sessions.
While a part of the publication (the medical file), can be encrypted using end-to-end encryption, some other parts must be used for routing the publication between authorized and interested parties.
The publication headers (name, address of the patient, nature of the test, etc.) are highly sensitive information.
Subscriptions are also highly sensitive information: they can reveal, for instance, which patient is treated by which clinic or for which type of ailment.
The leakage of such information can lead to serious consequences: one can imagine an insurance company observing such data and refusing to cover patients undergoing certain tests.
Furthermore, confidentiality management can get even more complex if the e-Health infrastructure interconnects with other systems through pub/sub communication, like a law enforcement agency gathering information about victims of a suspicious accident.
Again, we note that there is a compromise between the ability to route messages based on some information and potential leakages to unauthorized parties. 

A characteristic of both scenarios above is that the pub/sub service provider can be a third party belonging to an administrative domain different from the ones of publishers and subscribers.
This third party may not be trusted to access sensitive data.
Furthermore, the use of virtualization and the lack of control on resource placement and communication in public clouds might pose serious confidentiality threats.
For instance, research has shown that exploits at the hypervisor level~\cite{Ristenpart2009Hey-you-get-off,Somorovsky2011All-your-clouds} or at the CPU cache level~\cite{Liu_YGHL_15} when virtual machines are collocated can be used to gather private information from a virtual machine running on a public cloud.
Since the support infrastructure may be prone to attacks, confidentiality should be provided by design such that even unauthorized and accidental access to information manipulated by pub/sub brokers cannot cause critical data leaks.

\subsection{Survey motivation}
\label{sec.previousSurvey}

Several surveys have been written about various aspects in pub/sub systems, all having different aims than the present one.
\cite{Eugster2003The-many-faces-of-pu} positions the pub/sub paradigm with respect to other communication paradigms.
\cite{Baldoni2005Distributed-Eve} focuses on scalable event routing and its relation with the underlying overlay network.
\cite{Filho2005A-Survey-on-Ver} concentrates on software engineering with the purpose of achieving versatility in pub/sub middleware.
\cite{Liu2003Survey-of-Publish-Su} considers overlays topologies.
\cite{Martins2010Routing-Algorit} focuses on routing algorithms.
None of these surveys targets confidentiality in pub/sub systems.
There are a few contributions covering and classifying some security and confidentiality aspects of pub/sub systems~\cite{Wang2002Security-Issues,Raiciu2006Enabling-confid,Bacon2010Security-in-Mul}, which we cover in later sections of this article.

Confidentiality problems arising in pub/sub systems are acutely relevant, and it can be challenging to view the wide and disparate array of solutions 
 in proper context.
Hence, our first objective for this survey is to provide a clear and comprehensive overview of the state-of-the-art tools developed for this purpose.
The second objective is to describe and assess how providing confidentiality affects the core conception and performance of pub/sub systems.
Finally, our third objective is to point out the challenges that must be overcome before confidentiality can be efficiently supported in large-scale pub/sub systems, and more generally to shed light on the relevant research directions in this emerging and promising field.

\subsection{Organization}

The organization of this survey is as follows.
Section~\ref{sec:aspects} summarizes and classifies the various flavors of confidentiality in pub/sub systems considered in the literature.
To do so, we introduce a generic functional pub/sub system model encompassing most of the existing work.
Section~\ref{sec:models} surveys confidentiality-preserving pub/sub solutions based on security models and non-specific security tools, allowing the core matching operation to be performed only in trusted domains.
Section~\ref{sec:confidentiality} surveys solutions specific to pub/sub that provide the ability to perform matching operations on encrypted data in untrusted domains.
Finally, we discuss the current challenges and unexplored issues in Section~\ref{sec.conclu}.



\section{Confidentiality in the context of pub/sub}
\label{sec:aspects}

\emph{Confidentiality} is the property for a communication system to prevent the disclosure of sensitive information carried in the exchanged messages. 
Confidentiality in the context of pub/sub systems is approached in several ways, which we overview in this section.

In Subsection~\ref{sec:aspects:funcmodel}, we describe a generic system model for content-based pub/sub systems, that can be simply restrained to match the topic-based model.
This model is used as the unsecured basis for the pub/sub security solutions we review.
In Subsection~\ref{sec:aspects:modeltrust}, we discuss the importance and implication of trust assumptions over the domains and entities of a pub/sub system, and illustrate these using a seminal example in Subsection~\ref{sec:ehealth}.
In Subsection~\ref{sec:aspects:secprops} we propose a classification of pub/sub confidentiality properties.
Finally, in Subsection~\ref{sec:aspects:directions}, we distinguish the two main research directions towards providing confidentiality in pub/sub systems.

\subsection{System model}
\label{sec:aspects:funcmodel}

A typical content-based pub/sub system is composed of a broker infrastructure that provides routing services, and two sets of clients: \emph{publishers} that submit \emph{publications} to the system and \emph{subscribers} that submit \emph{subscriptions} with the intent to receive the publications that match their interests.
Brokers receive and store subscriptions from the subscribers that connect to them.
They also typically maintain routing tables towards other brokers, forming an overlay of brokers.
Upon the reception of a publication from a publisher or another broker, a broker checks its locally stored subscriptions and the entries in its routing table for matching interests (the \emph{matching} operation).
For each matching subscription, the broker sends a notification to the corresponding, locally attached subscriber.
For a matching routing table entry, the broker forwards the publication to the corresponding broker.
We assume without loss of generality that all brokers may perform the three operations of matching, notification and routing publications.
We consider in the following the general case of a content-based pub/sub system.
This can be adapted to a topic-based model simply by reducing the expressiveness of subscriptions to simple equality matching over a single attribute.

The structure of a publication includes a \emph{header} that defines the \emph{attributes} on which routing is based and their respective values (e.g., price = 300, name = ``\texttt{ACME}'', date = 2015/6/1), as well as a \emph{payload} that contains the complete data to be delivered (e.g., a graph showing the daily variations of the stock value).
Note that in practice a publication can be represented only by its header and the payload is optional.
Also, the number of attributes effectively present in a publication header can be lower than the total possible number in the publication schema which we refer as publication \emph{dimensions}.

The structure of a subscription consists in a set of \emph{constraints} on the attributes (e.g., price \texttt{>} 300 and name = ``\texttt{EMCA}'').
In this survey, when no distinction is necessary between subscriptions and publications (in particular for encryption purposes), an attribute from a publication header and a constraint from a subscription are both denoted as a message \emph{field}.

In order to efficiently match publications against large set of stored subscriptions, pub/sub systems often leverage \emph{containment} relations between subscriptions.
A subscription contains another subscription if it is more general, i.e., if publications matching the contained subscription always match the containing subscription (e.g., ``S$_1$: stockquote \texttt{>} 100'' contains ``S$_2$: stockquote \texttt{>} 300'').
Determining containment relations between subscriptions allows building efficient data structures to store subscriptions and match incoming publications.
An example of such a structure is a partially ordered set (or \emph{poset}), where precedence relations in the poset correspond to containment relations between subscriptions.
When a subscription is known not to match a publication, all its descendants in the poset can be marked as not matching as well (a process known as negative matching).
Similarly, a subscription known to match can mark all subscriptions that contain it as matching as well, by following containment order relations in the reverse direction (positive matching).
Containment is largely used by state-of-the art pub/sub systems~\cite{Li2005A-Unified-Appro,Jacobsen2009The-PADRES-Publ,Carzaniga2001Design-and-Evaluatio,Chand2004XNet:-a-Reliable-Con}.
It is shown in~\cite{Barazzutti2012Thrifty-Privacy} and other work that using containment may reduce the number of actual subscription matching evaluation by up to an order of magnitude compared to the naive one-by-one evaluation of the set of subscriptions, yielding important performance and scalability gains.

The majority of the work on confidentiality in content-based pub/sub systems follows the model described above, which fits naturally a deployment as an infrastructure service running in the cloud or on dedicated servers.
When one of the security schemes reviewed in this survey considers a slightly different system model, we explicit the difference in the text.
The system model for pub/sub that differs the most from a broker-based system is the use of peer-to-peer techniques, where publishers and subscribers self-organize in a broker-less overlay in order to perform the matching, routing and notification.
Examples include Meghdoot~\cite{Gupta2004Meghdoot:-content-ba} and Ferry~\cite{Zhu:2007:FPA:1263126.1263234} based on distributed hash tables, Sub-2-Sub~\cite{Voulgaris2006Sub-2-Sub:-Self} and~\cite{Costa2003Introducing-reliabil} using gossip-based protocols, and specific overlays such as R-trees~\cite{Bianchi2007Content-based-Publis}.
While there are less published solutions addressing confidentiality in peer-to-peer pub/sub, we include relevant work in this area in our review and explicitly detail the system model differences when applicable.

\subsection{Threat models}
\label{sec:aspects:modeltrust}

The starting point for research on confidentiality is the definition of a \emph{threat model} linked to the system model described above.
We identify in this subsection the commonalities and differences among the threat models found in the literature.

The behavior generally considered is that of \emph{honest-but-curious} entities.
Under this assumption, entities (publishers, subscribers, brokers) are considered untrusted to access the sensitive information but act according to the system specification (i.e., they are not malicious).
In particular, they will not replay, delete or forge messages.
However, they wish to collect any information allowing them to access restricted content in the exchanged messages, or equivalently provide this ability to external entities.
The published work we review define threat models that share the following traits: the system functionality (routing publications according to submitted subscriptions), the roles of the entities (subscribers, publishers and brokers), the data that is manipulated (publications and subscriptions) and the honest-but-curious behavior.
Differences in the threat models lie in the \emph{trust assumptions} that are made, and in the granularity with which these trust assumptions are set.
Trust assumptions explicitly separate entities forming the pub/sub system as trusted and untrusted for accessing specific pub/sub data (e.g., particular sets of fields in pub/sub messages).
The threat models of the work we review consider that untrusted entities do not collude in order to break confidentiality, e.g., by sharing the collected information in order to form a more powerful attack.

\subsection{Motivating example: extended e-Health infrastructure}
\label{sec:ehealth}

\begin{figure}[t]
\centering
\includegraphics[scale=0.5]{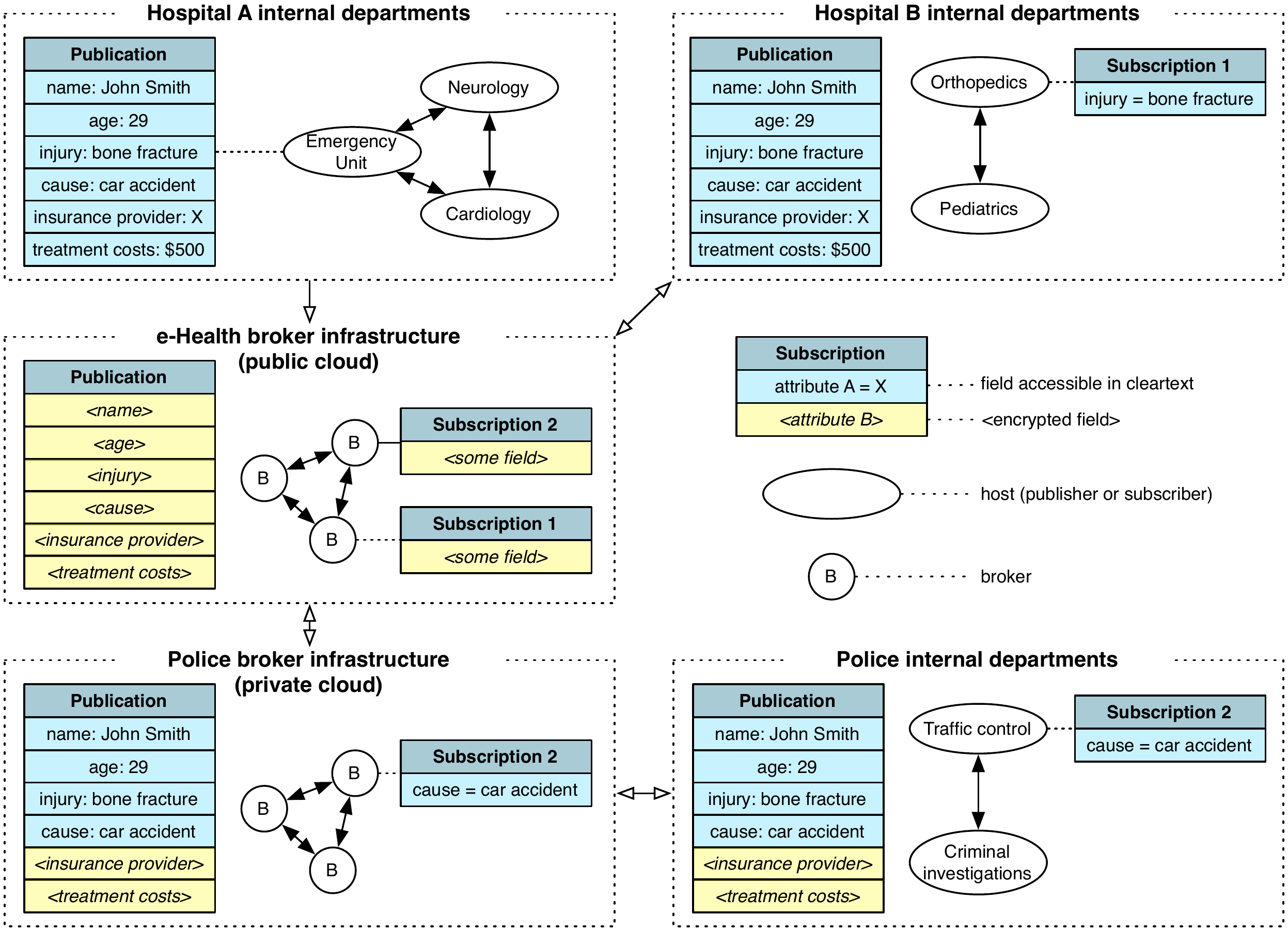}
\caption{E-health infrastructure: message flow traversing multiple domains with different trust assumptions for accessing the data.}
\label{fig.ehealth}
\end{figure}

Trust assumptions often derive from the potentially complex relationships between different administrative domains.
To illustrate this, we describe an extension of the e-Health use case introduced in Section~\ref{sec:intro:case}.
Figure~\ref{fig.ehealth} presents sample message flows and identifies, for each administrative domain, the pub/sub data that needs to be protected against untrusted nodes in that domain.
Each hospital is part of a separate domain.
It publishes patient files for new cases, and subscribes to take act of new cases from other hospitals.
Hospitals are trusted to access the publications and subscriptions in their local domain (e.g., when sending the file between two internal departments).
An e-Health pub/sub service implemented by a broker infrastructure allows communication between hospitals that are part of the countrywide public health system.
This infrastructure can be hosted on a public cloud; it is then untrusted for any access to pub/sub data.
In particular, this prohibits hospitals from sending their subscriptions constraints and publication headers in the clear to this domain for routing purposes.
A second broker infrastructure is used by police services for internal and external communication.
This infrastructure connects to the e-Health system in order to collect information such as the name, age and type of injury of victims involved in specific cases such as car accidents. 
A confidentiality policy requires however that some information from the patient files, for instance the cost of the treatment or the insurance provider, remains hidden to the police.
Unlike the e-Health system, the broker infrastructure of the police is not deployed on a public cloud but on privately owned servers.
Consequently, the brokers in this domain can be trusted to access the subscriptions constraints and some of the publication attributes, and use these in the clear for routing, although publication attributes that fall under the confidentiality policy must remain hidden.

\begin{figure}[t]
\centering
\includegraphics[scale=0.5]{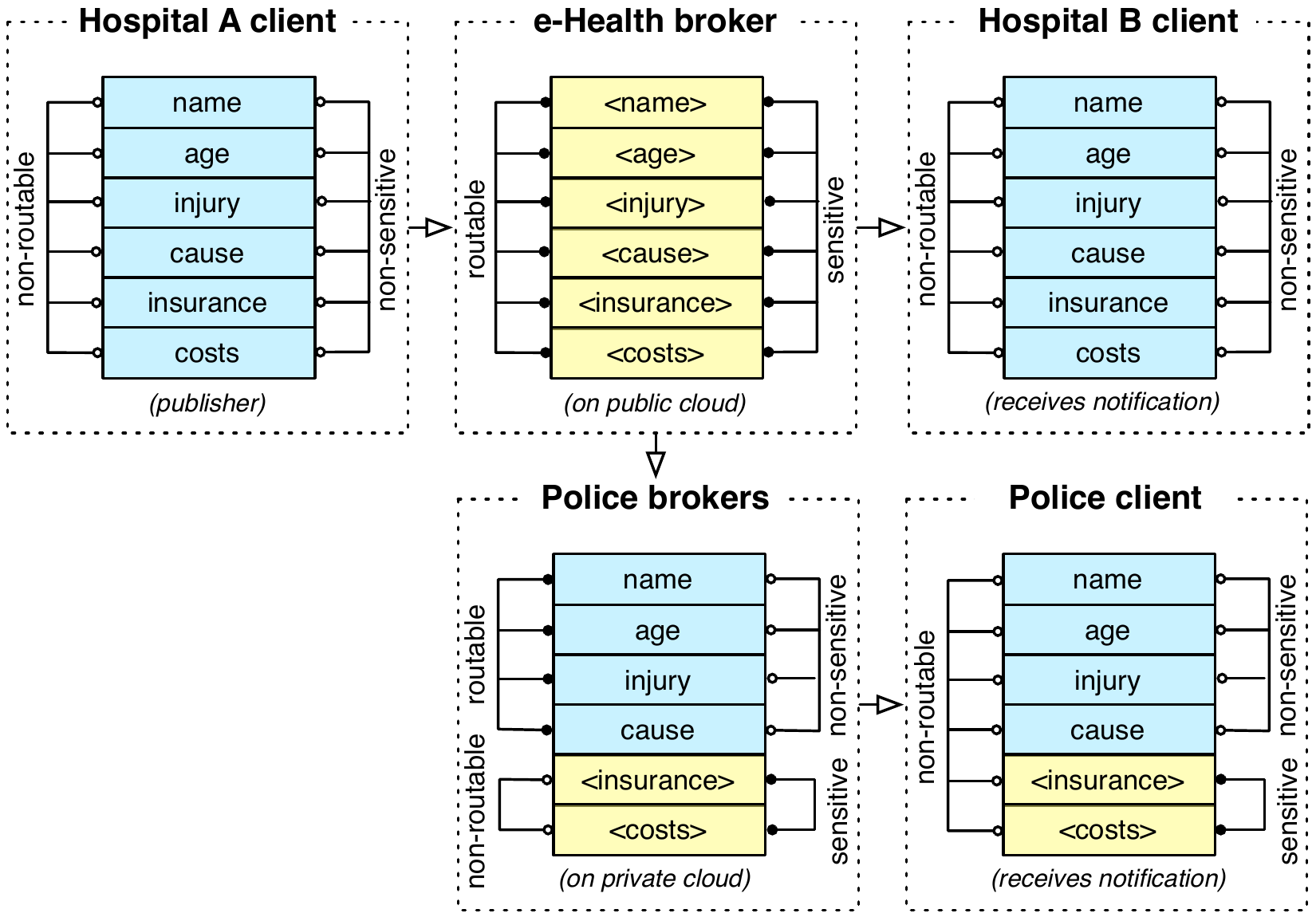}
\caption{Variation of publication field characteristics depending on the domains traversed.}
\label{fig.flow}
\end{figure}

This example illustrates the pairwise relationships between pub/sub entities in a given domain and the data that is exchanged.
On the one hand, there are functional requirements that are dictated by the role of the entity: brokers need to perform matching and routing operations between subscriptions and publications, whereas publishers and subscribers do not. 
On the other hand, trust assumptions dictate what pub/sub data should be available in each domain: entities in one domain may not be trusted to read any of the message content (e.g., the brokers of the e-Health system) but the same message can be partially accessed as it enters a different domain (e.g., the police broker infrastructure).
The content of publications and subscriptions can thus be classified as follows:
\begin{itemize}
\itemb \emph{routable} and \emph{non-routable} fields based on the necessity, in a specific domain, to use the content of the header fields for routing the messages;
\itemb \emph{sensitive} and \emph{non-sensitive} fields based on the need for protection against confidentiality threats according to the trust assumptions made in each domain (e.g., the name of the patient is a sensitive field in the e-Health system infrastructure, but it is non-sensitive in the police broker infrastructure).
\end{itemize}
We illustrate in Figure~\ref{fig.flow} how publication fields vary as a message passes through the various domains in the e-Health infrastructure. 
Note that the purpose of this example is to be simple but general enough to cover the assumptions and models found in the literature.
It can be generalized to more complex scenarios.
For instance, we considered that all pub/sub entities in a domain have the same role, a simplifying assumption that is generally used in the literature and does not break generality.

\subsection{Pub/sub confidentiality properties}
\label{sec:aspects:secprops}

We distinguish the three following confidentiality properties that apply to the different natures and roles of the sensitive parts of pub/sub messages:
\begin{itemize}
	\itemb \emph{Subscription confidentiality} is the ability to hide part or all of the constraints of the subscriptions when these subscriptions enter untrusted domains.
	\itemb \emph{Publication confidentiality} is the protection of part or all of the information contained in the \emph{headers} of publications, which are used for routing against stored subscriptions.
	\itemb \emph{Payload confidentiality} is the property of hiding the payload of the publications submitted to the pub/sub system.
\end{itemize}

A separation of confidentiality properties in pub/sub systems was first proposed in~\cite{Wang2002Security-Issues}.
We note the following differences with our classification.
First, Wang et al. consider our publication confidentiality property as two separate properties,  ``information confidentiality'' and ``publication confidentiality'', where information confidentiality protects the headers of publications when manipulated by the brokers, and publication confidentiality protects the entire publications against the subscribers that are not allowed to access them.
We argue that \emph{publication confidentiality} for headers can be provided against any untrusted system component, regardless of its nature.
The second difference between our classification and Wang et al.'s is that they do not consider payload confidentiality as a separate confidentiality property.
Since the payload is not used for routing purposes, enforcing payload confidentiality is orthogonal to the specifics of pub/sub communications.
The payload can thus be encrypted using classical encryption techniques before entering the untrusted domain and decrypted when re-entering a trusted domain, provided that the appropriate key distribution has been performed.

Subscription and publication confidentiality must be enforced through encryption of the corresponding sensitive parts.
Classical encryption can be used to enforce these two properties before pub/sub messages enter untrusted domains, however the brokers present in the untrusted domain lose the ability to perform routing operations on the sensitive parts.
Classical encryption is thus only applicable to non-routable fields. A class of encryption techniques specific for pub/sub allows encrypting publication headers and subscription constraints while preserving the ability to do content-based routing. 
These encryption mechanisms and the corresponding matching operation, which we denote as \emph{encrypted matching}, allow preserving the functional model for publications and subscriptions parts that are defined as both routable and sensitive in a given domain.

\subsection{The two lines of research on pub/sub confidentiality}
\label{sec:aspects:directions}

Published work on pub/sub confidentiality can be classified in two principal categories: research focused on leveraging existing security models and techniques, and research focused on providing pub/sub specific forms of encryption. While these two lines of research are not necessarily mutually exclusive, most of the existing work falls into one of the two categories, which makes the classification natural. 

\subsubsection*{Research focused on security models}

The first research avenue is geared towards finding optimal architectures and security models for pub/sub infrastructures that must respond to confidentiality threats.
The cryptography mechanisms used are ``conventional'': sensitive fields in messages must be encrypted before entering untrusted domains.
This encryption is opaque, and no operation can be performed on the encrypted fields.
In particular, it is not possible to have fields that are both sensitive and routable, as no matching decision can be made using their encrypted form.
The threat models are generally detailed and relatively complex: different trust assumptions can be set for the different domains and system components depending on the sensitivity of the different pub/sub message parts, allowing fine grain control.
For instance, in our extended e-Health scenario detailed in Section~\ref{sec:ehealth}, differences in trust assumptions between different infrastructures,  at the level of individual fields in subscriptions, can be expressed and enforced.
This is generally achieved by deriving access control matrices in which pub/sub entities are the subjects, pub/sub messages constitute the objects, and trust assumptions are formalized as access rights. 
We review the literature on this topic in Section~\ref{sec:models}.

\subsubsection*{Research focused on encrypted matching} 

This second line of research focuses on devising encryption techniques specific to pub/sub, and in the majority of cases \emph{content-based} pub/sub.
These encryption techniques allow matching encrypted subscriptions against encrypted publications, without requiring prior decryption and without revealing their content.
This is the only way to support fields that are both routable and sensitive for a domain containing untrusted brokers.
There is an important amount of work on devising such encrypted matching schemes, which we survey in Section~\ref{sec:confidentiality}.

\medskip

Each research direction suffers for a major drawback.
Existing work on security models precludes any routing on sensitive data in untrusted environments, which severely limits its potential applications.
Likewise, the body of work on encrypting matching often assumes simplistic threat models: every broker is untrusted and all the fields in pub/sub messages are sensitive when passing through the broker infrastructure.
We advocate bridging these two research avenues in our concluding notes in Section~\ref{sec.conclu}.


\section{Confidentiality in pub/sub security models}
\label{sec:models}

\begin{figure}[]
\centering
\includegraphics[width=1\textwidth]{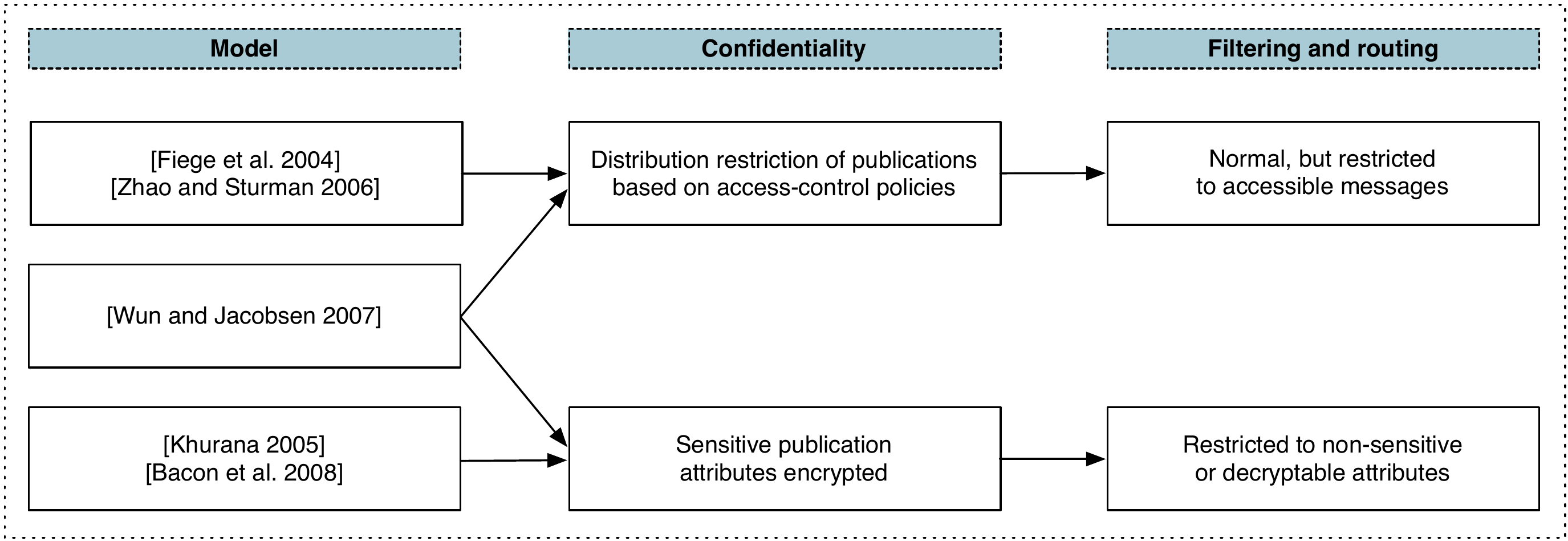}
\caption{Overview of pub/sub security models.}
\label{tab.sumtrust}
\end{figure}

The first line of research covered in our survey focuses on \emph{security models} for pub/sub infrastructures.
The solutions we overview in this section generally consider security models based on access control.
Different trust assumptions apply to the various domains to which the publishers, subscribers or brokers belong. Pub/sub data confidentiality must still be preserved, but the granularity of what data must be protected depends on the security model definition.
For instance, some of the brokers might be allowed to access part of the content of publications and subscriptions when they are granted the required authorization. 
An illustration is the routing infrastructure used by the police services in our e-Health scenario described in Section~\ref{sec:ehealth}.
The police brokers are allowed to access in cleartext and route against only a subset of the attributes in a publication's headers, but not allowed to access other attributes in the same headers. 

All solutions reviewed in this section target the \emph{content-based} pub/sub paradigm.
They are actually generally oblivious to the nature of the matching operation that is performed, as their primary goal is to ensure that only authorized entities get access to the fields used for routing \emph{in the clear}, after which they can perform an arbitrary type of matching operation.
This matching operation could be, without loss of generality, a single equality on a single attribute matching the topic-based model.

The rest of the section is organized as follows. We first review generic broker-based pub/sub systems in Subsection \ref{sec:generic}, followed by solutions used in ubiquitous computing in Subsection~\ref{sec:ubiquitous}. We conclude the section by discussing the limitations of the existing security models in Subsection~\ref{sec:modellimitations}.

\subsection{Generic broker-based pub/sub systems}
\label{sec:generic}

The work we overview in this subsection does not focus on the development of cryptographic mechanisms themselves but leverages existing general-purpose encryption schemes.
The solutions reviewed in this section and their main properties are summarized in Figure~\ref{tab.sumtrust}.

\medskip
\noindent
\textbf{\cite{Bacon2008Access-control-}} describes a pub/sub security solution that uses a role-based access control model (RBAC).
The initial form of this architecture is described in~\cite{Belokosztolszki2003Role-based-acce} and extended in~\cite{Bacon2005Securing-publis,Pesonen2005Secure-event-ty,PesonenAccessControl:2007,Pesonen2007Encryption-enfo}. 
It builds upon OASIS~\cite{BaconOASIS:2002}, an RBAC solution for distributed systems, and integrates the Hermes pub/sub platform~\cite{Pietzuch2002Hermes:-A-Distribute}.
Confidentiality is addressed in particular in~\cite{Bacon2008Access-control-}.
Additional work on security aspects in pub/sub systems that relate to confidentiality is detailed in~\cite{Bacon2010Security-in-Mul} for multi-domain systems and in~\cite{Singh:2011} for disclosure control.

\emph{Architecture.} 
\cite{Bacon2008Access-control-} considers that the participants to the pub/sub system (publishers, subscribers, brokers) span multiple administrative domains.
One domain is designated as the \emph{coordinating domain}, and its role is to invite other domains to join the pub/sub system.
Each domain includes an \emph{access control manager} responsible for enforcing rights for pub/sub operations for its own nodes. 
\emph{Key group managers} are in charge of administrating \emph{key groups} for rights related to cryptographic operations.
This operation is distinct from that of the access control managers.

\emph{Functionality.}
The functionality model is based on role based access control (RBAC).
Pub/sub system clients can be granted the right to define message types (e.g., the right to register a ``patient data'' publication type).
Clients become the owners of the message types they register and can define the associated access control policies.
A policy associates the actions each role is allowed to perform for the message type.
A role can be associated with the right to advertise a message before its publication, to publish and/or subscribe, but also to modify existing types or define new ones.
The enforcement of type policies is delegated to the access control managers in the administrative domains.

Cryptography is used to provide fine-grained access control and thus confidentiality enforcement in the system.
The initial encryptions and final decryptions of pub/sub messages before their deliveries are carried out by edge brokers (i.e., the brokers that are directly connected to the publishers and subscribers). 
An underlying assumption is that edge brokers are trusted to access the pub/sub data from the clients that belong to their own domain and thus connect to them.
Besides this assumption, access rights for non-edge brokers are set in the policies defined by type owners.
Similar to the functionality of access control managers, key group managers enforce the policies that authorize brokers to join key groups.
A key group consists of brokers that share the same level of access to the keys required to encrypt or decrypt pub/sub data.
\cite{PesonenAccessControl:2007} and~\cite{Bacon2008Access-control-} suggest using OFT (One-way Function Trees)~\cite{Sherman:2003} for key management in the groups that are formed.
\cite{Pesonen2007Encryption-enfo} mentions AES~\cite{DaemenR02} in EAX mode~\cite{Bellare:2003} as the preferred encryption scheme. 

\emph{Enforcing confidentiality.}
The level of protection for pub/sub data depends on the encryption granularity.
Two cases are considered: encryption of the whole message and encryption \emph{per attribute}. 
The first case allows any broker to route messages based on their type, which is always accessible.
This is similar to topic-based routing.
Authorized brokers can decrypt complete messages and perform content based routing. 
When encryption per attribute is used, an independent encryption key is associated to each of the attributes. The routing capacity of each broker depends on the attributes it can decrypt according to its given rights: if a broker can (cannot) decrypt an attribute, then it can (cannot) route based on the content of this attribute. This allows to model fine-grained trust assumptions by the type owners but also implies a level of overhead for key management that grows proportionally with the number of accessible attributes.

\medskip
\noindent
\textbf{\cite{Zhao:2006}} presents a service model for providing access control in a pub/sub system.
This work focuses on the ability to change the access rules at runtime.
The ideas are implemented in the Gryphon~\cite{Strom1998Gryphon:-An-Inf} pub/sub system.

\emph{Architecture.}
Access to pub/sub data going through multiple administrative domains is regulated through access control and a versioning control is used for changing the access control policies.
A \emph{security administrator} entity has the responsibility of adding or performing changes to the rights granted to the pub/sub system clients.
Subscribers and publishers are not trusted to perform actions on pub/sub data unless specifically authorized.

\emph{Functionality.} 
Pub/sub clients obtain rights for certain actions such as connect, publish or subscribe if they authenticate successfully as a \emph{principal}.
This is similar to a role: multiple clients can run on behalf of the same principal.
The service model formalizes access control rules into rights using an extended format for pub/sub messages: [\emph{Principal, Type of Access, Filter}].
The type of access can represent simple actions (e.g., connecting to the system) or complex actions (e.g., registering a subscription).
The access right is given by the filter, which for simple actions can be a boolean value (e.g., granting or not the connection right) and for complex actions an extended filter (e.g., setting restrictions for allowed subscription or publication types).
The rights are stored in a database by the security administrator, which also handles rights changes (the version control). 
Such changes are processed in atomic batches distributed across the broker network. 
The brokers synchronize with the security administrator on the starting point, in the message data stream, where the new rights apply.
This synchronization guarantees the consistency of the access control rules throughout the whole system.

\emph{Enforcing confidentiality.} 
Confidentiality for publications is achieved by limiting their distribution to subscribers through subscriptions rights.
Also, the authors mention that brokers with different levels of trust should be partitioned into  different domains, regulating communication through access control policies. 

\medskip
\noindent
\textbf{\cite{Khurana2005Scalable-securi}} presents a security model relying on proxy re-encryption, targeting a pub/sub system for secure XML document dissemination~\cite{Bertino:2002}. 
The scheme is based on~\cite{Jakobsson:1999}.

\emph{Architecture.}
The threat model assumes that brokers are not trusted to access sensitive parts of pub/sub messages, and routing is only performed using non-sensitive fields.
The system requires the presence of trusted servers that host a proxy security and accounting service (PSAS).

\emph{Functionality.} 
Only the sensitive and non-routable parts of publications are encrypted.
Dissemination of the publications relies on secure XML document distribution~\cite{Bertino:2002}, which itself relies on symmetric encryption.
A publication encryption key is further encrypted with a public key belonging to the publisher and sent along with the encrypted publication.
PSAS servers are in charge of distributing the symmetric publication encryption key.
The PSAS also acts upon broker requests by transforming the messages encrypted with the public key of a publisher into messages encrypted with the public keys of the subscribers.
This avoids a direct exchange of keys between the source and destinations, but requires that a quorum of PSAS servers be constantly available to ensure service availability.
Consequently, these PSAS servers  constitute single points of failure and may impair the scalability of the solution.

\emph{Enforcing confidentiality.} 
Confidentiality is considered only for publications, and solely enforced for the non-routable attributes.
The authors do not specifically state how publication payloads should be protected, but we assume  that these can be encrypted using the key used for encrypting non-routable attributes.

\medskip
\noindent
\textbf{\cite{Fiege2004Security-Aspect}} presents a security solution based on scopes, which allow defining groups of entities with the same trust level and authorizations.
The solution is implemented in the Rebeca pub/sub system~\cite{Muhl:2010}.

\emph{Architecture.} 
Publishers, subscribers and brokers are joined in different groups named \emph{scopes}.
The communication within a scope is isolated.
The admission of new members to a scope requires credential authentication according to the scope acceptance criteria.

\emph{Functionality.} 
The definition of scopes depends on common trust relationships established between different administrative domains.
A scope can recursively be a member of other scopes.
Scopes are used to limit the visibility of messages to their members through the use of access control policies.
The enforcement of these policies is based on checks performed on attribute certificates~\cite{Farrell:RFC3281}.
These represent credentials consisting of a signed identity associated to a set of attributes that belong to the entities in the system.
Attribute certificates are issued either by an attribute authority or by the owner of the broker network that belongs to a certain scope. 

Nodes obtain credentials and use them when advertising a type of publication (the functional model requires this step before publishing) or when they try to submit subscriptions.
Edge brokers check credentials and may allow the dissemination inside the scope.
A broker tries to match the attributes in a subscriber's certificate to the ones in the publisher's advertisements that was previously received.
If this check fails, the subscriptions will not be processed.

\emph{Enforcing confidentiality.}  
The confidentiality property is not specifically addressed in the paper.
However, the access control mechanisms provide confidentiality against any node that falls outside a scope of trust. This does not modify the content of messages for protection, but merely restricts their dissemination.

\medskip
\noindent
\textbf{\cite{Wun2007A-Policy-Manage}} presents a policy management framework offering general services for pub/sub architectures. 
It goes beyond security enforcement, which is nonetheless presented as the main use case. 
The framework was implemented on top of the PADRES~\cite{Jacobsen2009The-PADRES-Publ} pub/sub system.

\emph{Architecture and functionality.} 
The design is centered on a \emph{post-matching} policy model. 
A policy contains condition-based rules that define actions to execute when the rules are triggered.
Rules can be triggered while or immediately after matching the publications with the subscriptions.
The main purpose is to avoid testing the rules \emph{a priori}, which would duplicate the matching functionality when using policy rules that are semantically based on the publication content. 

The article considers a simple setting of \emph{trust groups}.
The entities forming the trust groups (publisher, subscriber, broker) interact as discussed next.
For security enforcement, the authors assume as a precondition that each trust group is associated with a shared group secret.
This secret information is used by pub/sub entities for authentication and encryption using specific group related protocols. 
The policy management framework preserves confidentiality using \emph{authenticated event scopes} and \emph{security zones}.

\emph{Enforcing confidentiality.}
Authenticated event scopes are associations of an authentication policy with rules in the post-matching model. 
For instance, if a message must be routed to a certain broker after it was matched, an authentication of the receiver might first be performed according to the policy.
Security zones extend the functionality of authentication policies by providing publication confidentiality at the granularity of attributes. 
The attributes in the matched publications can be either pruned or encrypted before forwarding  based on criteria given by established zones with different policy settings.

\subsection{Pub/sub privacy for ubiquitous computing}
\label{sec:ubiquitous}

The pub/sub dissemination model can be applied to sensor networks used in ubiquitous computing.
Using our e-Health example from Section~\ref{sec:ehealth}, consider an array of sensors for home patients monitoring temperature, heart rate, etc. 
These sensors periodically report measurements on which the system can react by notifying a physician based on registered subscriptions (e.g., $temperature > 37.5$).

Due to power and computational constraints in mobile environments, the existing work does not consider complex confidentiality preserving schemes.
The approach is limited to simple access control mechanisms, which is conceptually close to the work surveyed in this section. 
We briefly present in the following two relevant examples, that both target the content-based model.

\medskip
\noindent
\textbf{\cite{Opyrchal:2006}} proposes an access control policy model sharing similarities with~\cite{Belokosztolszki2003Role-based-acce} and~\cite{Bacon2008Access-control-}.
An application consists of multiple event types, and the administrator of an application is an entity who can grant the right to perform actions within the application, including adding new event types and delegating ownership of events.
An event owner is an entity who has the right to authorize actions over an event type, including publishing, subscribing, receiving and changing the policy of an event.
These rights can be further delegated to other licensed entities. 
The brokers enforce the access control and can restrict the registration of subscriptions or the delivery of publications according to the policy settings, including dynamic conditions (e.g., limiting the number of events delivered to a subscriber). 

The test case is a location tracking application: users carry RFID badges, and sensors detecting their location report the data to location publishers who then send the data to the brokers.
The architecture is thus similar to a generic enterprise broker-based pub/sub system.

\medskip
\noindent
\textbf{\cite{Tian:2013}} 
considers a centralized pub/sub system formed of a privacy engine, a subscription manager, and a matching engine having the role of a broker.
Sensor data is captured by device agents and transmitted to the pub/sub system.
The privacy engine attaches privacy points to the subscriptions constraints and publication attributes.
For instance, in our e-Health monitoring example, a subscription could be \{$\mathit{temperature} < 37,5$, $t < 5$\}, where $t$ is a privacy threshold.
A temperature measurement reported to the system can have a privacy point based, for instance, on its location, like \{$\mathit{temperature} = 38$, $t = 4$\} if the patient is in the kitchen or \{$\mathit{temperature} = 38$, $t = 6$\} if the patient is in the bedroom. 
Functions for privacy points can make the system more flexible: event delivery could be allowed when the patient temperature reaches a critical point but denied otherwise.
We remark again that the access control design is orthogonal to the sensor network itself and thus similar to a generic enterprise broker-based pub/sub system.

\subsection{Limitations of surveyed security models}
\label{sec:modellimitations}

Despite allowing fine-grain trust relations and control over the sensitivity of fields in pub/sub messages, the work surveyed in this section has a major limitation: it does not allow brokers in untrusted domains to route publications using sensitive fields in said domains. 
In our e-Health example of Figure~\ref{fig.ehealth}, the e-Health broker infrastructure hosted in a public cloud would not be able to filter patients information flowing from Hospital A to Hospital B or the police infrastructure. It could only forward all messages sent between these three domains. The system will ultimately have to rely on one of the two following solutions:
either all messages have to be flooded and filtered in the destination domain, where the required routable fields are non-sensitive, or all subscriptions must be replicated from the subscriber domains to all the publishers domains in order to filter messages at the source. These bypass solutions imply either a waste of bandwidth and a poor scalability, or increased management complexity, loss of the decoupling between producers and consumers of information, and storage overhead. 

As a direct consequence, the brokers' inability to perform computations over sensitive fields, precludes also the possibility to perform additional optimizations at the broker level, such as leveraging subscription containment. 
Although determining subscription containment can be seen as a confidentiality liability in some cases~\cite{Barazzutti2012Thrifty-Privacy,Barazzutti:2015aa,Raiciu2006Enabling-confid}, depending on the level of security desired in the pub/sub application domain, it can also consist in a viable tool for improving performance.

These fundamental limitations greatly decrease the appeal of using public cloud infrastructures to host pub/sub routing services, despite its numerous advantages like availability, cost-effectiveness and elasticity~\cite{Barazzutti:2014db}.
This limitation led to the development of a new class of \emph{encrypted matching} schemes, allowing routing over sensitive fields.
We review the existing encrypted matching schemes in the next section.


\section{Confidentiality through encrypted matching}
\label{sec:confidentiality}

In this section, we survey the existing work on encrypted matching, a form of encryption that allows routing based on the content of sensitive, thus encrypted data.
We first start by describing the general characteristics of encrypted matching schemes in Section~\ref{subsec.genencmatch}, followed by the presentation of the schemes themselves in Section~\ref{subsec.solencmatch}.

\subsection{General properties of encrypted matching schemes}
\label{subsec.genencmatch}

Encrypted matching schemes are used in pub/sub systems to encrypt subscriptions and publication headers.
We define the functional requirement of encrypted matching algorithms as solely the ability to perform the matching operation between a subscription and a publication, where at least one of the two messages is encrypted.\footnote{The majority of the schemes encrypt both publications and subscriptions, but this relaxed definition allows us to include exceptions.}
The algorithms do not need to decrypt as long as this matching operation is possible.
The flow of subscriptions ends at one of the brokers, where they are stored in their encrypted form, whereas the flow of publications ends at the subscribers, who are usually interested in the publication payload (if the header information is also of interest, it can be added in the payload).

We split the matching schemes based on the algorithm class they fit into: \emph{private-key cryptography} and \emph{public-key cryptography}.
In \emph{private-key cryptography} (also known as symmetric-key), the same private key information is used for both encryption and decryption, whereas in \emph{public-key cryptography} (asymmetric) a public key is used for encryption and a different private key is used for decryption.
We are aware that this classification is somewhat misleading due to the lack of decryption, but it remains the most natural in this context. 
While there was no published work providing a finer classification and formal model for encrypted matching classification, we suggest and argue in Section~\ref{sec:modelsconclusion} that \emph{functional encryption}~\cite{functional_encryption_waters} would be a good starting point for defining such a model, a work that is however beyond the scope of the survey aspect of the current paper.

\subsubsection{Matching algorithms} 
\label{matchalg}

Published work can be classified in two categories based on the encrypted matching operation:

\paragraph*{Matching based on an exact relation preserving isomorphism}
Consider a function $\Diamond$ applied on a publication attribute $a$ and a constraint value $c$  such that a matching relation between $a$ and $c$ can be determined.
For instance, the difference $\Diamond(a,c) = a-c$ can determine a ``greater than'' relation and whether or not the attribute matches the constraint. 
Consider now the encryption algorithm $E$ and a function $\Box$ applied on the ciphertexts $E(a)$ and $E(c)$. If an isomorphism 
	\begin{equation}
	  \label{eq:1}
	  \Diamond(a,c) = \Box(E(a),E(c))
	\end{equation}
can be established, then based on the result of $\Box$ we can determine the matching between $a$ and $c$. 
A typical example is the asymmetric scalar product-preserving encryption scheme (ASPE)~\cite{Choi2010A-Privacy-Enhan}.

Matching schemes based on isomorphisms include schemes based on \emph{homomorphic encryption}.
Such schemes have the property that given two plaintext terms $a$ and $c$ and two operations $\Diamond$ and $\Box$, we have
	\begin{equation}
	  \label{eq:2}
	   E(\Diamond(a, c)) = \Box(E(a), E(c)). 
	\end{equation}
The difference between (\ref{eq:1}) and (\ref{eq:2}) is that in (\ref{eq:2}) the function $\Diamond$ is encrypted.
Providing $\Diamond$ to a broker is thus insufficient to determine the relationship between $a$ and $c$ without decrypting, however an untrusted broker cannot be allowed to decrypt messages.
To overcome this issue, classic homomorphic schemes must be adapted for encrypted matching scenarios, as done in~\cite{Nabeel2009Privacy-Preserv,Nabeel:2012}.

\paragraph*{Matching based on a pre-mapped equality comparison}
The publication attributes and subscription constraints are pre-mapped to sets of values that permit matching strictly based on equality comparisons.
An encryption scheme $E$ is then applied to the pre-mapped results so that the equality can still be evaluated over the resulting ciphertexts.
The pre-mapping operation varies. In some cases, approximations are used to handle inequality constraints. Other instances consider the bit prefixes of constraint $c$ and attribute $a$; 
if a prefix in the set associated to $a$ is equal to a prefix from the set associated to $c$, then the matching is positive.

Matching schemes using pre-mapped equality comparisons include schemes that rely on \emph{attribute-based encryption (ABE)}, although in most cases is used for access control and not for encrypted matching. 
In some cases (e.g.,~\cite{Tariq:2014}), the scheme design does not require or consider preserving both subscription and publication confidentiality, and the ABE decryption also serves implicitly for encrypted matching.
Generally, an extra encryption layer must be added besides ABE to hide the attributes (publication headers) and the access structure (subscription constraints), and to perform encrypted matching over these.
Pre-mapping to bit prefixes in conjunction with ABE was used in~\cite{Ion:2012,Ion2010Supporting-Publ}.

It can be argued that matching based on a pre-mapped equality comparison is a special case of matching based on a relation preserving isomorphism.
However, it is natural to distinguish both approaches since there are cryptographic schemes supporting only equality comparisons with a preliminary mapping phase.

\subsubsection{Types of constraints handled}

The constraints of a subscription can be classified based on two criteria:
\begin{itemize}
	\itemb \emph{type of field} - numerical or string;
	\itemb \emph{type of operator} - equality or range for numerical types, and identity, prefix, suffix and substring for strings. 
\end{itemize}
In our description of the encrypted matching algorithms, we pay special attention to schemes that allow inequality matching since they provide the most subscription expressiveness and thus are the most interesting for content-based routing.
Conversely, a scheme supporting only equality operators and a single dimension falls into the simpler topic-based model.

\subsubsection{Matching performance}

Encrypted matching schemes evaluate a function over an encrypted constraint and an encrypted publication attribute.
Existing schemes however differ greatly in terms of the supported constraints and matching technique, and comparing the performance of the various solutions is cumbersome. 
It is nevertheless possible to determine when the encrypted matching between a single constraint and an attribute requires computations over a larger set of values (e.g., the complete set of publication attributes).
This can add a significant computation load, which makes these schemes unusable for workloads with numerous attributes per publication.
We point out such cases in our overview.

\subsubsection{Cryptanalytic attack models}

A cryptographic scheme has to withstand a range of cryptanalysis attacks whose purpose is to obtain the plaintext or the encryption key.
There are four \emph{basic} attack models considered when evaluating the security of classic cryptographic schemes.
They are based on the power of the attacker:

\begin{itemize}
	\itemb \emph{Ciphertext-only attack (COA)} when the attacker can only observe the ciphertexts;
	\itemb \emph{Known-plaintext attack (KPA)} when the attacker can observe the exact correspondence between ciphertexts and plaintexts; 
	\itemb \emph{Chosen-plaintext attack (CPA)} when the attacker has the power to obtain the ciphertexts corresponding to plaintexts of its choice;
	\itemb \emph{Chosen-ciphertext attack (CCA)} when the attacker has the power to obtain the decryption of ciphertexts of its choice.
\end{itemize}

Security analyses of CPAs or CCAs are usually formalized as an indistinguishability game, where the attacker must distinguish ciphertexts with a non-negligible probability.
Unfortunately, due to the need to perform the matching operation, encrypted matching for pub/sub is different from typical encrypted communication. 
We therefore concur with the conclusions in~\cite{Raiciu2006Enabling-confid} that the level of indistinguishability that is achievable is limited in this setting.
By matching encrypted publications against encrypted subscriptions, an untrusted broker can infer similarities between publications.
This is even more problematic for schemes that can leverage subscription containment, because the ability to determine that all events matching a subscription $s_a$ will match a containing subscription $s_b$ allows an attacker to infer similarities between the subscriptions themselves, raising the effectiveness of the attacker.
For this reason, the security analysis of existing encrypted matching schemes does not usually rely on indistinguishability proofs and the \emph{chosen} attack models CPA and CCA are seldom considered.
Instead, existing work either defers to the weaker KPA or COA or defines other attack models.
A common assumption for the behavior of brokers is to consider them as honest-but-curious.

\subsubsection{Key management} 

The main goals of key management are to:
\begin{itemize}
	\itemb allow system components to obtain or negotiate in a secure manner the initial information (keys or other parameters) necessary to establish a secure communication;
	\itemb allow key information to be refreshed at the different parties when necessary: when a member is evicted from the system and is not trusted anymore, when a member joins the system and is not trusted to decrypt previous communicated information, to counter brute force attacks, etc.
\end{itemize}

All encrypted matching schemes we overview require a form of exchange or negotiation of secret information between the participating system components, even for solutions relying on public key algorithms.
Pub/sub systems, and in particular encrypted matching schemes, introduce important challenges for key management.
First, the pub/sub decoupled communication model makes it cumbersome to identify subscribers and publishers who might need common key information.
Second, encrypted subscriptions stored by brokers are invalidated by a key refresh; 
this can cause service disruption. 
None of the encrypted matching schemes we overview proposes or references a key management alternative that addresses both of these issues:

\begin{itemize}
	\itemb In most cases, key management is simply considered an orthogonal problem decoupled from the cryptographic mechanism, and is required from an external entity.
	\itemb For schemes relying on attribute-based encryption, key information is associated with the content of the pub/sub messages rather than with particular system parties. 
	This attempts to reduce the coupling required by the key exchange.  
	\itemb  The pub/sub functional model is modified towards a less decoupled model, therefore allowing more straightforward designs for key management. 
\end{itemize}
Key management is an open and challenging issue for pub/sub systems, far from being satisfactorily solved by current research.
For this reason, challenges and directions towards potential solutions are discussed in details in Section~\ref{sec.conclu}.

\subsection{Overview of encrypted matching solutions}
\label{subsec.solencmatch}

\begin{floatingtable}[hl]{
	\scriptsize
	\begin{tabular}{cc}
	\toprule
		\textbf{Symbol} & \textbf{Role} \\
	\midrule
		$S$ & subscription \\
		$P$ & publication \\
		$\mathcal{A}$ & publication attributes \\
		$\mathcal{|A|}$ & number of attributes \\
		$\mathcal{C}$ & subscription constraints \\
		$\mathcal{|C|}$ & number of constraints \\
		$d$ & number of dimensions \\
		$m$ & bit size of values representation\\ 
		$E$ & encryption algorithm \\ 
          	$D$ & decryption algorithm \\
          $E_K$ & encryption algorithm with key $K$ \\ 
          $D_K$ & decryption algorithm with key $K$ \\
	\bottomrule
	\end{tabular}
	}
\caption{Notations.}
\label{tab.notations}
\end{floatingtable}

In the following we overview existing pub/sub encrypted matching schemes.
Many of these schemes use complex mechanisms, sometimes partially changed over different published versions of the same scheme.
We delve into particular mechanism details only when necessary for a better understanding of the scheme. 
Therefore, we often use a different, simplified notation compared to the original articles. 
We write $E(x)$ for a ciphertext obtained using an encryption algorithm $E$ applied to the plaintext $x$, and $D(y)$ for a decryption algorithm $D$ applied to a ciphertext $y$.
We write $E_{K}(x)$ and $D_{K}(y)$ when using a particular key $K$. The additional notation we use in this section is summarized in Table~\ref{tab.notations}.

\begin{figure}[]
\centering
\includegraphics[width=1\textwidth]{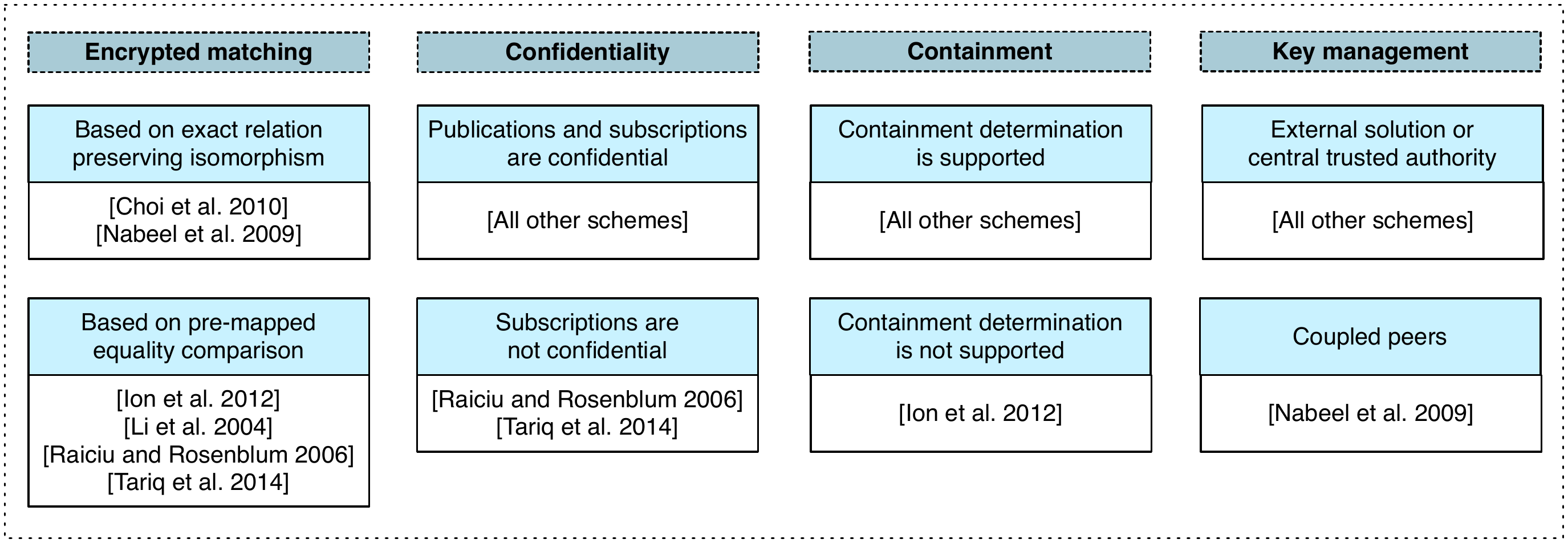}
\caption{Overview of representative encrypted matching solutions.}
\label{tab.sum}
\end{figure}

Figure~\ref{tab.sum} summarizes the characteristics of the most representative solutions surveyed based on the characteristics described in Section~\ref{subsec.genencmatch}: type of encrypted matching, confidentiality, containment support, and key management.
Table~\ref{tab.complexities} compares the representation of publications and subscriptions before they are encrypted.
This is relevant since it has an important impact in practical implementations.
For instance not all schemes permit encryption of publications that do not include field values for all the dimensions.
Table~\ref{tab.complexities} also specifies the estimated computational complexities for the message encryption and encrypted matching operations.

\paragraph*{Supported pub/sub paradigms}
Most of the solutions surveyed in this section target the \emph{content-based} pub/sub paradigm.
Their support for equality constraints (in some cases with reduced complexity compared to more general types of constraints as seen in Table~\ref{tab.complexities}) make them suitable for a topic-based installation as well.
Two exceptions are the schemes presented in~\cite{Song:2000} and~\cite{Crescenzo:2013}, that only support encrypted matching for equality constraints.
The principles of these two schemes are actually similar, and the same principles are used as a basis for more complex schemes such as the one in~\cite{Raiciu2006Enabling-confid} (supporting the content-based model through a combination of techniques for different types of constraints) or~\cite{EURECOM+2633} (supporting only equality constraints but on \emph{multiple} attributes, hence classified in the content-based paradigm).
The EventGuard system~\cite{Srivatsa:2011} also only supports equality matchings when protecting publication and subscription confidentiality using encrypted matching, but the main focus of this system is on the protection of payload confidentiality.
Protecting payload confidentiality only obviously allows using an arbitrary filtering model using attributes and constraints in the clear.


\begin{table}[t]
\setlength\tabcolsep{1pt}
\scriptsize
\centering

\newcolumntype{E}{ >{\centering\arraybackslash} m{0.5\linewidth} }
\newcolumntype{I}{ >{\centering\arraybackslash} m{0.39\linewidth} }

\newcolumntype{F}{ >{\arraybackslash} m{0.08\linewidth} } 
\newcolumntype{G}{ >{\centering\arraybackslash} m{0.25\linewidth} }
\newcolumntype{J}{ >{\centering\arraybackslash} m{0.13\linewidth} }
\newcolumntype{H}{ >{\arraybackslash} m{0.8\linewidth} }
\newcolumntype{K}{ >{\centering\arraybackslash} m{0.8\linewidth} }

\begin{tabular}{FGGJJJ}
\toprule
	&
	\multicolumn{2}{E}{\textbf{Representation}} &
	\multicolumn{3}{I}{\textbf{Complexity}} \\
	&
	Publication & Subscription &
	Enc. (Pub) & Enc. (Sub) & Match \\
\cmidrule{2-6}

	\multicolumn{6}{H}{\textbf{\cite{Choi2010A-Privacy-Enhan} }} \\

	\multicolumn{6}{H}{\hspace{5mm} \textbf{$\rightarrow$ constraints $<, \geq, >, \leq, =$ }} \\
	
	&
	1 point in $d$ dims. &
	$|\mathcal{C}|$ points in $d$ dims. &
	$O(d^2)$ &
	$O(|\mathcal{C}|\cdot d^2)$ &
	$O(|\mathcal{C}| \cdot d)$ \\
	
\cmidrule{2-6}

	\multicolumn{6}{H}{\textbf{\cite{Raiciu2006Enabling-confid} }} \\

	\multicolumn{6}{H}{\hspace{5mm} \textbf{$\rightarrow$ equality constraint $=$ }} \\	

	&
	$|\mathcal{A}|$ (dim., val.) pairs &
	$|\mathcal{C}|$ (dim., val.) pairs &
	$O(|\mathcal{A}|)$ &
	$O(|\mathcal{C}|)$ &
	$O(|\mathcal{C}|)$ \\
	
	\multicolumn{6}{H}{\hspace{5mm} \textbf{$\rightarrow$ conjunctive set membership and $<, \geq, >, \leq$ }} \\	
	
	&
	\multicolumn{2}{E}{bit field \emph{dictionary} of size $D$} &
	$O(D)$ &
	$O(|S|)$ &
	$O(|S|)$ \\
	
\cmidrule{2-6}

	\multicolumn{6}{H}{\textbf{\cite{Ion2010An-implementati,Ion2010Supporting-Publ,Ion:2012} }} \\
	
	\multicolumn{6}{H}{\hspace{5mm} \textbf{$\rightarrow$ equality constraint $=$ }} \\
	
	&
	$|\mathcal{A}|$ (dim., val.) pairs &
	$|\mathcal{C}|$ (dim., val.) pairs &
	$O(|\mathcal{A}|)$ &
	$O(|\mathcal{C}|)$ &
	$O(|\mathcal{C}|)$ \\
	
	\multicolumn{6}{H}{\hspace{5mm} \textbf{$\rightarrow$ constraints $<, \geq, >, \leq, =$ }} \\
	
	&
	$|\mathcal{A}|\cdot m$ tokens, each with 1 bit set&
	$|\mathcal{C}|\cdot O(m)$ tokens, each with 1 bit set&
	$O(|\mathcal{A}|\cdot m)$ &
	$O(|\mathcal{C}|\cdot m)$ &
	$O(|\mathcal{C}|\cdot m^2)$	\\

\cmidrule{2-6}

	\multicolumn{6}{H}{\textbf{\cite{Nabeel2009Privacy-Preserv,Nabeel:2012,Nabeel:2013} }} \\

	\multicolumn{6}{H}{\hspace{5mm} \textbf{$\rightarrow$ constraints $<, \geq, >, \leq, =$ }} \\
	
	&
	$|\mathcal{A}|$ (dim., val.) pairs &
	$|\mathcal{C}|$ (dim., val.) pairs &
	$O(|\mathcal{A}|\cdot \log n)$ &
	$O(|\mathcal{C}|\cdot \log n)$ &
	$O(|\mathcal{C}|)$ \\

\cmidrule{2-6}

	\multicolumn{6}{H}{\textbf{\cite{Li2004An-efficient-sc} }} \\

	\multicolumn{6}{H}{\hspace{5mm} \textbf{$\rightarrow$ constraints $<, \geq, >, \leq, =$ }} \\

	&
	$|\mathcal{A}|$ (dim., val.) pairs&
	$|\mathcal{C}|$ (dim., $I$) pairs&
	$O(|\mathcal{A}|\cdot m^2)$ &
	$O(|\mathcal{C}| \cdot m^3)$ &
	$O(|\mathcal{C}|\cdot m)$ \\
	
	
\cmidrule{2-6}

	\multicolumn{6}{H}{\textbf{\cite{Tariq:2010,Tariq:2014} }} \\

	\multicolumn{6}{H}{\hspace{5mm} \textbf{$\rightarrow$ constraints $<, \geq, >, \leq, =$ }} \\

	&
	all $L$ domain decompositions containing pub. point &
	smallest domain decomposition containing sub. range(s) &
	$O(L)$ &
	\emph{(does not apply)} &
	$O(d)$ \\

\bottomrule

\end{tabular}

\caption{Computational complexity and pre-encryption data representation of encrypted matching schemes. For~\protect\cite{Choi2010A-Privacy-Enhan}, we have $|\mathcal{A}|=d$ as all attributes must have a value in a publication. For \protect\cite{Raiciu2006Enabling-confid}, $|S|$ is the number of items in the subscription set to be tested for inclusion in the publications' sets. For~\protect\cite{Nabeel2009Privacy-Preserv} $n$ is the upper limit of the Paillier cryptosystem plaintext domain. For~\protect\cite{Li2004An-efficient-sc} $I$ is the set of prefixes covering the range of a constraint and whose size is $O(m)$.} 
\label{tab.complexities}
\end{table}

\medskip
\noindent
\textbf{\cite{Choi2010A-Privacy-Enhan}} describes a solution using \emph{asymmetric scalar product-preserving encryption} (ASPE).
ASPE was initially introduced by~\cite{Wong:2009} for secure kNN query computation on encrypted databases. The solution applies to any numerical constraint.

\begin{figure}[h]
\centering
\includegraphics[scale=0.5]{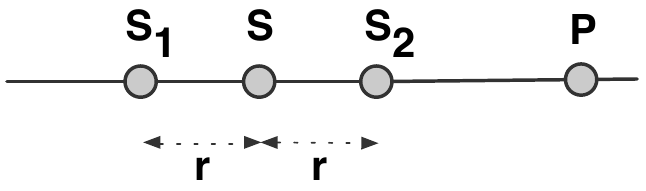}
\caption{The ASPE subscription split.}
\label{ASPEsplit}
\end{figure}

\emph{Scheme mechanism.}
Publication attributes and subscriptions constraints are represented as coordinates of multidimensional points. 
The scheme relies on computations and comparisons that use the Euclidean distance between points in the multidimensional space.
The simple 1-dimension case is shown in Figure~\ref{ASPEsplit}.
Consider a subscription point $S$ and a publication point $P$ that must be matched. Without loss of generality, suppose that we want to find whether $S < P$.\footnote{The cases $S\leq P$, $S>P$, $S\geq P$ and $S=P$ are done similarly.}
The subscriber chooses two reference subscription points $S_1$ and $S_2$ located symmetrically at equal but random distance $r$ from $S$.
The distance difference $D_1 = d(S_1, P) - d(S_2, P)$ can then be compared to zero to determine the relation between $S$ and $P$: 
$$d(S_1, P) - d(S_2, P) > 0 \Leftrightarrow d(S_1, P) > d(S_2, P) \Leftrightarrow S < P.$$ The distance difference $D_1$ can be expressed as a sum of scalar products:
$$D_1 = d(S_1,P) - d(S_2,P) = \parallel S_1 \parallel ^2 - \parallel S_2 \parallel ^2 + 2(S_2-S_1)\cdot P.$$

The core principle of ASPE is to allow the encryption of points $S_1$, $S_2$ and $P$ while preserving the ability to compute the scalar product.
This allows the algorithm to determine the distance difference $D_1$ but does not preserve the actual distance between $S$ and $P$.
The motivation behind this strategy is that \cite{Wong:2009} proves that any encryption scheme preserving the distance between two points is vulnerable to known-plaintext attacks (KPA).

The key on the subscriber side is an invertible matrix $M$, whereas on the publisher side the key is $M^{-1}$. The matching phase relies on the key reduction when the ciphertext encrypted with $M$ is multiplied with the one encrypted with $M^{-1}$. 
The result obtained is $D_2=D_1 q$, i.e., the distance difference $D_1$ multiplied by a positive random scalar $q$.
The scheme is thus based on an exact relation preserving isomorphism, as discussed in Section~\ref{matchalg}. Subscription containment can be supported by adding extra reference points to the subscription. 

\emph{Security considerations.} 
The security evaluation in~\cite{Choi2010A-Privacy-Enhan} is rather shallow and only considers a particular ciphertext-only attack (COA). 
To guarantee subscription and publication confidentiality, the authors rely on the more comprehensive security proofs given in the original database scenario~\cite{Wong:2009}. 
However, to strengthen the security, \cite{Wong:2009} enhance the original scheme by splitting the original dimensions and introducing additional artificial dimensions.
These additions are not discussed in~\cite{Choi2010A-Privacy-Enhan}, but they appear to be adaptable to the pub/sub case. 

\emph{Practical aspects.}
ASPE is a private-key algorithm, which requires that the key information be distributed among participants. 
However, the paper does not give any information about key management. 
Another limitation is the multidimensional case (multiple constraints and attributes) that is also only briefly discussed. 
Our own analysis on this indicates that the encrypted matching depends on the size $d$ of the schema defining the set of publication attributes, the complexity being $O(d)$ per evaluated constraint. This yields a quadratic complexity for matching one subscription if the number of constraints is close to the number of dimensions. 
Therefore, the solution is less appropriate for workloads that require a large number of attributes.

We can consider as a practical use case the stock quote notification system referenced in Section~\ref{sec:intro:case}. 
The set of dimensions for the published data in such a scenario is not very large: a symbol name and various fields related to the quote value and variation.
Although ASPE supports only numerical constraints, it can accommodate string fields like symbol name or stock market indicator by simply mapping their values to a numerical domain.
This is possible when the operation used on such fields is limited to equality comparisons.
ASPE supports ranged evaluation over encrypted data, which can be particularly useful in pub/sub workload such as the monitoring of gain or loss trends in stock quote variations.
This scenario involves periods of high throughput publications and a large number of active subscriptions depending on the stock market activity, which are well suited for optimizations based on the containment support of ASPE. However, these optimizations decrease the confidentiality of the scheme~\cite{Barazzutti:2015aa}.

\medskip
\noindent
\textbf{\cite{Raiciu2006Enabling-confid}} presents a set of encrypted matching mechanisms that use different encryption techniques according to the type of values (integer or string) and the nature of subscriptions constraints (equality or range).

\begin{figure}
	\centering
	\includegraphics[scale=0.5]{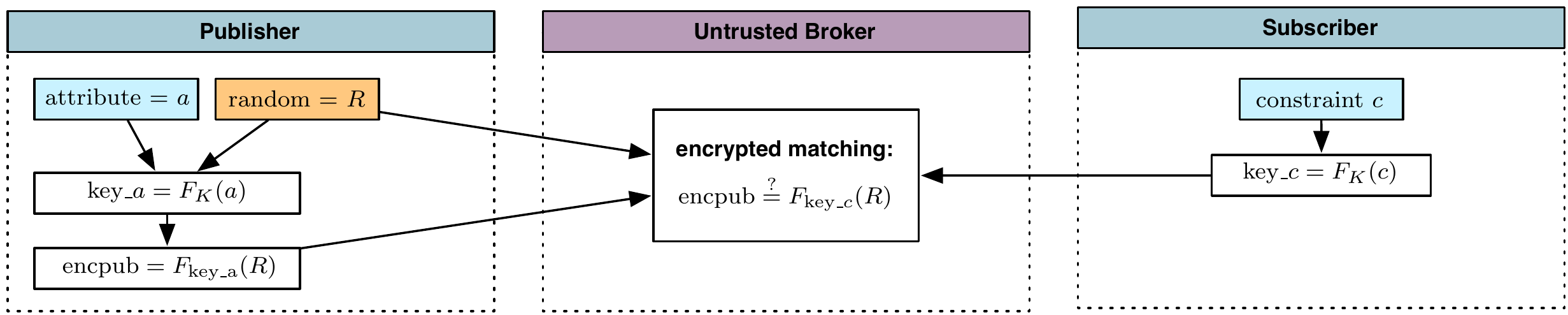}
	\caption{Encrypted matching for equalities in~\protect\cite{Raiciu2006Enabling-confid}.}
	\label{RReq}
\end{figure}

\emph{Scheme mechanism.}
For equality filtering, the paper uses a simple scheme initially proposed by~\cite{Song:2000} and used  in~\cite{EURECOM+2633,Srivatsa2007Secure-Event-Di}.
The mechanism is depicted in Figure~\ref{RReq}.
Subscribers and publishers share a common secret $K$.
A parameterized pseudorandom function $F_{k}$ is applied to the publication attribute $a$ at the publisher, and to the constraint $c$ at the subscriber. 
While $\text{key\_}c = F_{K}(c)$ is sent directly to the broker, $\text{key\_}a = F_{K}(a)$ is used as the parameter for another application of $F$, this time to a random token $R$.
This yields the encrypted publication $\text{encpub}$.
$R$ and $\text{encpub}$ are sent to the broker, who is able to compute $F_{\text{key\_}c}(R)$.
If this is equal to $\text{encpub}$ (denoted by the $\stackrel{?}{=}$ sign) then the equality constraint matches.
In summary, the scheme idea is using the ciphertexts of the attributes and the constraints as keys, while the actual encrypted plaintext compared in the matching phase is random.
If the keys are equal then the matching is positive.

This simple scheme is extended to conjunctive set membership.
The application proposed by the authors is to check for the presence of a set of keywords in the subscription in a set of keywords representing the publication.
The mechanism can be applied to any set membership problem where the definition domain of the set is known and of small size, and can be easily modified to support disjunctive set membership.
We present an example of this extension in the following.

\begin{figure}
	\centering
	\includegraphics[scale=0.5]{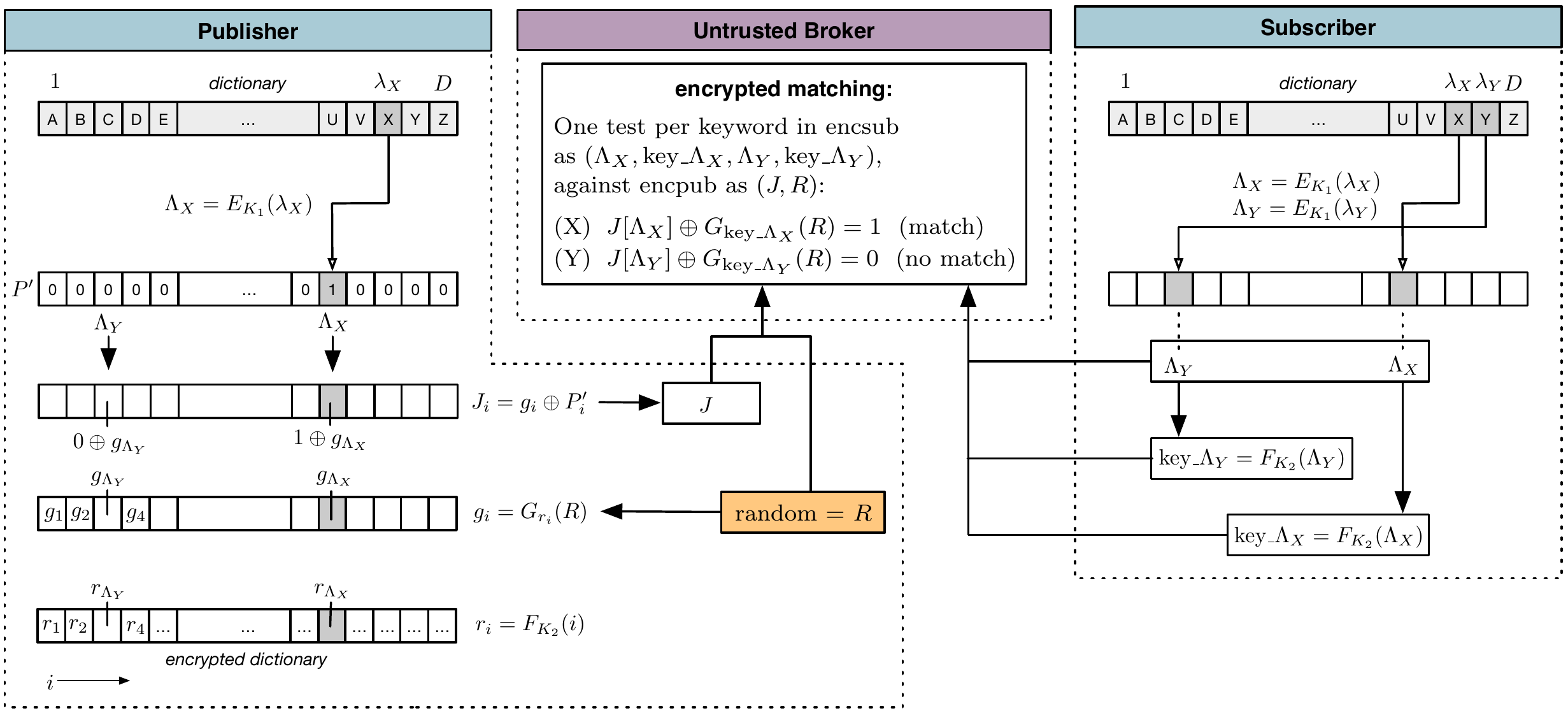}
	\caption{Dictionary-based encrypted matching for conjunctive set membership in~\protect\cite{Raiciu2006Enabling-confid}.}
	\label{RRscheme_example}
\end{figure}

The algorithm considers a dictionary of size $D$ that indexes the $D$ possible values in the domain, and a randomly chosen secret key $K$ formed of two halves $\{K_1,K_2\}$.
Both are available to all publishers and subscribers.
Since the dictionary must index all elements from the set's domain, this domain must be of bounded and known size---this size must also remain reasonably small as the space complexity of encrypted publications is in O($D$).

Figure~\ref{RRscheme_example} illustrates the matching of a subscription $S$ with conjunctive constraints for items $Y$ and $Z$ against a publication $P$ including only the item $Y$.

$P$ and $S$ can be modeled as the indexes $\lambda$ of their items in the dictionary, i.e., $P=\{\lambda_X,\lambda_Y\}$ and $S=\{\lambda_X\}$ ($|P|$ and $|S|$ denote the size of $P$ and $S$, respectively). The first step of the encryption for both the publication and the subscription is to permute these $\lambda$ indexes using a first shared keyed pseudorandom permutation function $E_{K_1}$.
This results in permuted indexes $\Lambda \in [1\dots d]$, e.g., $\Lambda_X=E_{K_1}(\lambda_X)$.
Theses indexes $\Lambda$ are used similarly to attribute $a$ and constraint $c$ in the equality matching scheme.

On the subscription side, $F$ is used with parameter $K_2$ to encrypt $\Lambda_X$ and $\Lambda_Y$, resulting in $\text{key\_}{\Lambda_X}=F_{K_2}(\Lambda_X)$ and $\text{key\_}{\Lambda_Y}=F_{K_2}(\Lambda_Y)$.
The encrypted subscription $\text{encsub}$ formed of the two pairs ($\Lambda_X,\text{key\_}{\Lambda_X},\Lambda_Y,\text{key\_}{\Lambda_Y}$) is sent to the untrusted broker.

On the publication side, the encryption uses three additional steps.
First, the scheme applies $F_{K_2}$ to all index positions $i \in [1\dots D]$ in the dictionary (not only the ones present in the publication).
This yields $D$ values $r_i=F_{K_2}(i)$, which are used as keys for another pseudorandom function $G$ applied to a random value $R$.
This results in a vector of encrypted dictionary indexes $g_i, i\in [1\dots D]$ where $g_i = G_{r_i}(R)$.
Finally, in order to differentiate the indexes of attributes present in the publication from the ones that are not, values $g_i$ corresponding to the permuted indexes $\Lambda$ of the publication are XOR'd (operation $\oplus$) with a vector of 1s.
The resulting vector is denoted $J$.
The encrypted publication $\text{encpub}$ sent to the untrusted broker is the pair ($J,R$).
If $J_{\Lambda_X} \oplus G_{\text{key\_}{\Lambda_X}}(R) = 1$, keyword $X$ appears in the publication.
If all constraints of the subscription are matched, the publication matches the subscription.

For inequality constraints, a variant of the set membership encrypted matching scheme can be applied through the use of an appropriate dictionary.
Numerical inequalities can be supported by choosing a set of reference points and ordering guarantees as the dictionary, e.g., $\{<1, <3, <5, >1, >3, >5\}$. 
In this case, an initial approximation of the values is required before encryption to pick the closest reference point in the dictionary.
The scheme thus follows a pre-mapped equality comparison design as defined in Section~\ref{subsec.genencmatch}. Subscription containment is determined in a similar manner by adding an additional reference vector with approximation points for each constraint.

\emph{Security considerations.}
Two identical subscriptions $S_1$ and $S_2$ will be encrypted as the same $\text{encsub}$, and the ability to determine containment relations leads to inherent distinguishability (even if containment support is optional).
The authors thus admit that in such a case, subscription encryption is not semantically secure and therefore focus on the broker's ability to distinguish between publications.
The analysis shows that the only information leaked is the positive or negative matching result.

\emph{Practical aspects.} 
All schemes considered in the paper are based on symmetric-key encryption, and require prior key (and dictionary) exchange between participants. No solution is propose to enable this exchange. 

For equality constraints, the scheme requires the individual encryption of each of the $|\mathcal{C}|$ (dimension, value) constraints in the subscription and one corresponding comparison operation in the matching phase.
The matching equality complexity is therefore in O($|\mathcal{C}|$).

For conjunctive set membership, the space complexity for a subscription is in O($|S|$), two values per item in the set.
Supporting containment requires an additional vector of size $D$ for each item, raising subscription space complexity to O($|S|\times D$).
Subscription encryption complexity is also in O($|S|$).
Publication encryption requires operations on all the elements of $D$-sized vectors, irrespective  of the number of attributes: space and time complexities are thus in O($D$).
Finally, matching consists of $|S|$ operations on single elements from the encrypted publication vector $J$ and random $R$ and is therefore in $O(|S|)$.

For inequalities, the complexity of the subscriptions is unchanged.
On average half ($\frac{D}{2}$) of the dictionary entries will be set for the publications, but as the space and encryption complexity for publications was already $O(D)$, it remains identical.

While a small dictionary size $D$ is recommended for performance purposes for set membership subscriptions, we observe that for inequalities $D$ corresponds to a cost-accuracy tradeoff: a small dictionary size will result in many false positives and increased traffic of unwanted publications towards subscribers, whereas a large value of $D$ increases costs.

We finally observe that the approximation phase used for the containment tests can also produce false negatives.
While this does not affect the matching of individual subscriptions, this can decrease the effectiveness of optimizations based on containment relations between subscriptions: an actual containment relation between two subscriptions might not be discovered, preventing from using positive matching as discussed in Section~\ref{sec:aspects:funcmodel}.

\medskip
\noindent
\textbf{\cite{Ion2010An-implementati,Ion2010Supporting-Publ,Ion:2012}} present an encrypted matching scheme that uses \emph{attribute-based encryption} (ABE) and multi-user \emph{searchable data encryption} (SDE). The scheme applies to any numerical constraint and to string equalities.

\begin{figure}
	\centering
	\includegraphics[scale=0.5]{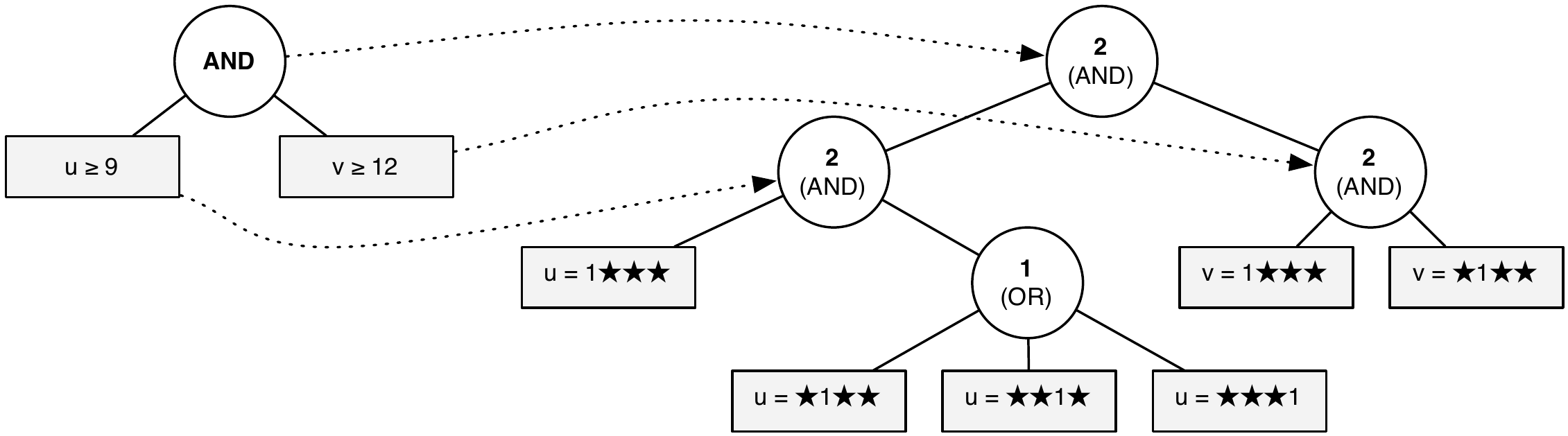}
	\caption{Subscription representation after attribute-based encryption (ABE) using four bits.}
	\label{fig.iontree}
\end{figure}

\emph{Scheme mechanism.}
We discussed the generalities of ABE, a form of functional encryption, in Section~\ref{subsec.genencmatch}. 
In this scheme, ABE is used for encrypting the publication payload. 
Publications can be decrypted only if an access policy associated with the subscriber that receives them is satisfied.
The access policy sets a specific structure for the subscription constraints and dictates the format of publications' attributes. 
Constraints are encoded into an access tree in which non-leaf nodes are threshold gates specifying the number of subtrees that need to be satisfied. 
As an illustration, consider the subscription with two constraints $\{u \geq 9, v \geq 12\}$
shown in Figure~\ref{fig.iontree}.
This subscription results in a root node with a value of 2, meaning that both subtrees (one per constraint) must be satisfied.
The constraints are expanded following the representation in~\cite{Bethencourt:2007}.
At most one leaf-node token per bit is required in order to make a decision.
For the predicate  ``$v \ge 12$'' using a 4-bit representation, the constraint subtree will have a leaf-node with $v=\texttt{1***}$ (``bit 1 of \emph{v} must be 1'') and a leaf-node with $v=\texttt{*1**}$ (``bit 2 of \emph{v} must be 1'') joined by a parent ``\texttt{AND}'' node. 
Publication attributes are also split into tokens with 1 bit set (e.g., $v=12$ is split as $\{v=\texttt{1***}$, $v=\texttt{*1**}$, $v=\texttt{**0*}$, $v=\texttt{***0}\}$). 
The matching scheme follows the generic pre-mapped equality comparison design we defined in Section~\ref{subsec.genencmatch}.
We now detail how the publication's tokens and the similarly structured subscription leaf nodes are encrypted in order for the matching to be possible over ciphertexts.

\begin{figure}[h]
	\centering
	\includegraphics[scale = 0.5]{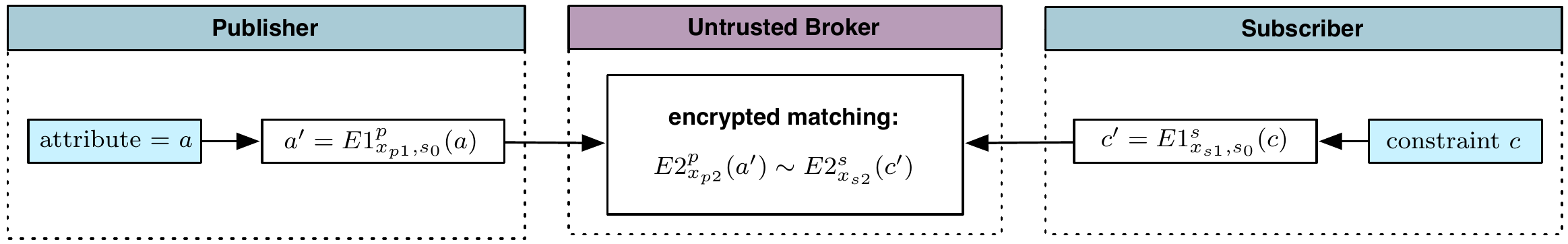}
	\caption{SDE private key usage applied in pub/sub context 
(~{\raise.17ex\hbox{$\scriptstyle\sim$}} denotes the matching operation at the broker)}
	\label{fig.ionmatch}
\end{figure}

The tokens with 1 bit set obtained from publications and subscriptions are encrypted using multi-user SDE~\cite{Dong:2008}. 
This technique adapts the public key El Gamal scheme~\cite{ElGamal:1985} to a proxy re-encryption context.
A trusted authority generates the public key parameters and a private key formed by two components $x$ and $s_0$.
The private part $s_0$ is known to both the publisher and subscriber, but not to the broker. 
The central idea of the scheme is to split the private key $x$ in two pieces 
for each pair formed by an end node (publisher or subscriber) and the edge broker through which that end node sends and receives messages.  
The sketch in Figure~\ref{fig.ionmatch} summarizes the technique.
The private key $x$ used for encryption by both the publisher and subscriber is split in two parts for each of these and the broker between them: $x = x_{p1}+x_{p2}$ where $x_{p1}$ is given to the publisher and $x_{p2}$ to the broker, and $x= x_{s1}+x_{s2}$ where $x_{s1}$ is given to the subscriber and $x_{s2}$ to the broker. 
We can generalize the mechanism by saying that for each of the communicating end nodes $i$ connected to a broker, the common key $x$ is split into $x_{i1}$ which is given to the end node and $x_{i2}$ which is given to the broker.
The end node performs an initial encryption $E1$ in which his private part $x_{i1}$ and $s_0$ are used. 
The broker performs a re-encryption $E2$ using his part of the private secret $x_{i2}$.
The purpose of splitting the same key $x$ in different pairs of $x_{i1}$ and $x_{i2}$ is the application of the proxy re-encryption mechanism.
We abstract the details of the effective algorithms used in encryption, which differ for publications and subscriptions.
The main idea of the scheme is that each end node-broker pair can share the same key $x$. 
Only the splits of $x$ for the end node and the broker differ.
Therefore, the ciphertexts obtained in the two steps (initial encryption and re-encryption) are comparable, being finally encrypted with the same key material, and the matching can be determined. 
However, subscribers use a different random parameter in the encryption of each subscription. 
This does not impact the matching but prevents brokers from determining containment relations between subscriptions ciphertexts. 

\emph{Security considerations.}
Subscription confidentiality is analyzed using a chosen plaintext attack (CPA) model; 
for publication confidentiality the scheme is evaluated using a model similar to the one in~\cite{Raiciu2006Enabling-confid}, the goal being that nothing is leaked to an adversary besides the results observed from matching history traces.
The analysis uses the fact that the employed El Gamal mechanism is proven CPA secure. 
Finally, the payload confidentiality provided by ABE is proven secure under a stronger threat model that takes into account the possibility of collusion between subscribers, publishers and brokers.

\emph{Practical aspects.}
Besides encrypted matching, the paper also describes an access control model that can be applied to more complex scenarios like the e-Health use case detailed in Section~\ref{sec:ehealth}. 
The access control model relies on restrictions that can be imposed on various groups of subscribers through ABE.
This fits with a multi-domain context such as the one in the e-Health use case.
Also, using ABE for publication payload encryption could be an appropriate solution for typical additional data included in the messages that might require selective authorized decryption rights (e.g., publications can include blood test analysis, X-rays, etc. accessible to only part of the medical personnel determined through a particular ABE access policy). 
The paper does not discuss the support of containment determination.
However, in an e-Health scenario the system load is expected to be relatively light compared to the stock exchange use case, and therefore the lack of containment optimization is less critical. 

The scheme was integrated to the PADRES~\cite{Jacobsen2009The-PADRES-Publ} pub/sub middleware for its evaluation, and tested in an e-Health context. 
\cite{Ion:2012} reports that the matching time increases linearly with the number of comparisons done between constraint and attribute tokens.
A constraint subtree can have at most $m$ leaf-nodes, and an attribute token set has $m$ elements where $m$ is the number of bits used to represent a value.
The worst-case scenario is that each attribute token with 1 bit set must be compared with each leaf-node token with 1 bit set, thus the maximum number of comparisons per constraint is $m^2$ in the general case.
This results in a matching complexity in $O(|\mathcal{C}|\cdot m^2)$ comparisons per subscription.
If subscriptions include only equality constraints, these do not have to be expanded to leaf node tokens with 1 bit set, and neither do the publication attributes: 
encrypted tokens can be represented directly by the constraint and attribute values, which results in only one comparison per constraint.
The general cost is, however, higher than other schemes using similar data representation (i.e., \cite{Li2004An-efficient-sc}), although the privacy guarantees are also stronger.

The article considers the lack of need for key exchange as a decoupling advantage.
However, this does not eliminate the need for common key material between communicating peers.
The difference is that this key material is not exchanged between participants, but is instead retrieved from a trusted authority. 

\medskip
\noindent
\textbf{\cite{Nabeel2009Privacy-Preserv,Nabeel:2012,Nabeel:2013}} present a solution applicable for any constraints on numerical values based on the public-key Paillier cryptosystem~\cite{Paillier:1999}.
The Paillier cryptosystem has the property that $E(x)\cdot E(y)=E(x+y)$, and the private key uses two components $\lambda$ and $\mu$ in the decryption phase.

\begin{figure}
\centering
\includegraphics[scale=0.5]{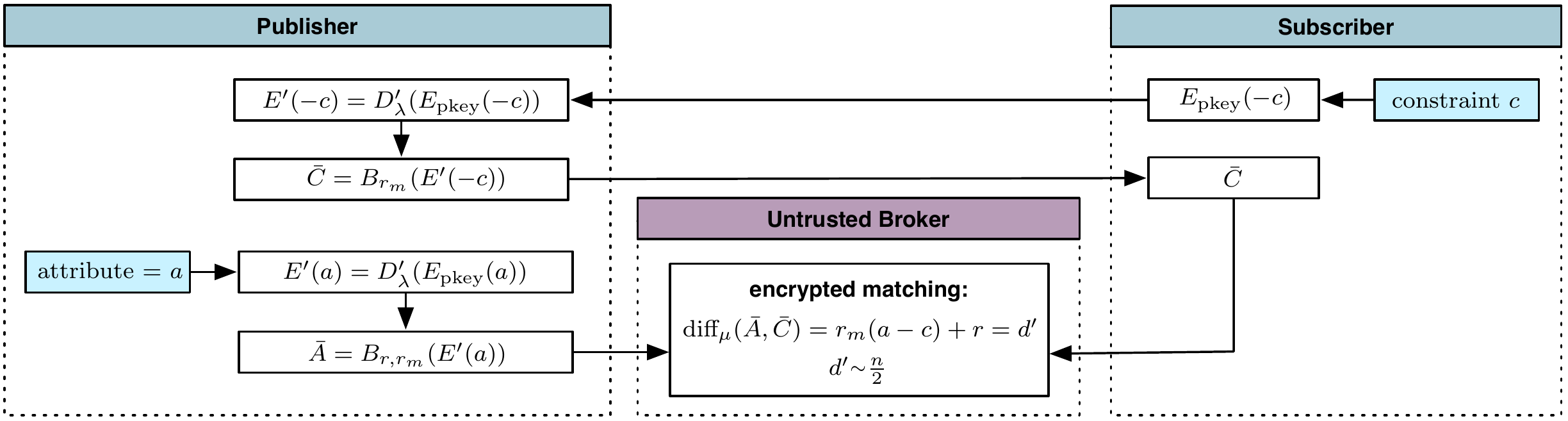}
\caption{Encrypted matching mechanism in~\protect\cite{Nabeel:2012} ($\text{pkey}$ and $\mu$ are public parameters, $\lambda, r$ and $r_m$ are private parameters, $\bar{A}$ and $\bar{C}$ respectively denote \emph{blinded} attributes and constraints values, $n$ is the size of the Paillier plaintext space, and ${\raise.17ex\hbox{$\scriptstyle\sim$}}$ is the matching comparison operation).
}
\label{Nabeel}
\end{figure}

\emph{Scheme mechanism.}
In the first two versions of the scheme~\cite{Nabeel2009Privacy-Preserv,Nabeel:2012}, subscribers know in advance the publishers from whom they wish to receive publications, and publishers are also aware of this intention. 
A sketch summarizing the encrypted matching is shown in Figure~\ref{Nabeel}.
The scheme follows the exact relation preserving isomorphism design described in Section~\ref{matchalg} through the use of the homomorphic properties of the Paillier cryptosystem.

In an initial phase, the subscriber encrypts the negation of each constraint value $c$ using its own public key $pkey$, obtaining $E_{pkey}(-c)$.
These values are sent to the publisher from which the subscriber wants to receive the publications.
The publisher then applies an additional blinding layer on top of the received Paillier encryption:
\begin{itemize}
	\itemb It applies a part of the Paillier decryption operation that consists in an exponentiation using the secret $\lambda$ parameter, which we denote $D'$ and whose result is $E'(-c)$: $D'_{\lambda}(E_{pkey}(-c)) = E_{pkey}(-c)^{\lambda} = E'(-c)$;
	\itemb It applies an additional encryption $B$ with a private random component $r_{m}$ to obtain the blinded value $B_{r_{m}}(E'(-c))$.
\end{itemize}
Blinded encrypted constraints are sent back to the subscriber, who registers the encrypted subscriptions with the broker. 
The publisher uses a similar technique to encrypt and blind each publication attribute $a$ using the same separate private $r_{m}$ and an extra random parameter $r$ to derive $B_{r,r_{m}}(E'(a))$.
For matching, the broker is given the second part of the Paillier private key $\mu$. 
Using $\mu$ and the homomorphic properties of the Paillier cryptosystem, the broker is able to perform a computation $\text{\emph{diff}}$ over the blinded attribute and constraint.
This computation completes the Paillier decryption and results in 
$$\text{\emph{diff}}_{\mu}\left(B_{r,r_{m}}(E'(a)),B_{r_{m}}(E'(-c))\right) = r_{m}(a-c)+r = d'.$$
The scheme assumes that the domain for the attributes and constraints values ($a$ and $c$) can be approximated to the interval $[0,2^l]$, where $2^l \ll n$, $n$ being the upper limit of the plaintext space of the Paillier cryptosystem.
Since the domain used in the pub/sub scenario is much smaller than the plaintext space, the difference result $a-c$ can be mapped in the mutually exclusive intervals $[0,2^l]$ if $a \ge c$ and $[n-2^l, n]$ if $a < c$.
The random values $r_{m}$ and $r$ in the $\emph{diff}$ result, chosen in controlled range, obfuscate the actual $a-c$ difference and permit expanding the mapping intervals to the complete plaintext space.
If $d' \le n/2$ then the broker can conclude that $a \ge c$, respectively $a < c$ if $d' > n/2$. 

Containment is optionally supported through a similar technique.
Another private parameter $r_{c}$ can be used by the subscriber to obtain separate blindings for the constraints values in a subscription, and their negations (e.g., $B_{r,r_{c}}(E'(c))$, $B_{r_{c}}(E'(-c))$). 
This allows the broker to compute differences as \emph{diff} above for two constraints values $c_1$, $c_2$ blinded in this manner:
$$\text{\emph{diff}}_{1}(B_{r_1,r_{c}}(E'(c_1)),B_{r_{c}}(E'(-c_2))) = r_{c}(c_1-c_2)+r_1;$$
$$\text{\emph{diff}}_{2}(B_{r_2,r_{c}}(E'(c_2)),B_{r_{c}}(E'(-c_1))) = r_{c}(c_2-c_1)+r_2.$$
 (where $r_1$ and $r_2$ are random values). 
The broker can derive containment based on these results. An aspect to consider is that the same value $r_{c}$ must be used in blinding the constraints values in different subscriptions.
Unless an agreement on the value can be established between different subscribers, containment can be leveraged only between subscriptions of the same subscriber.

The latest variation of the scheme~\cite{Nabeel:2013} introduces a context manager. 
It acts as a trusted entity, is responsible for managing contexts represented by sets of attributes in publications, and distributes the security parameters required for encryption and blinding in a certain context.
The encrypted matching computation performed on the brokers remains essentially the same as presented previously.
However, \cite{Nabeel:2013} does not mention explicitly containment support.
In this version of the scheme a subscriber no longer communicates directly with the publisher for the initial blinding phase of the subscription.
Subscribers receive the security parameters necessary for obtaining the blinded result from the context manager. 
In addition to the encrypted matching support based on the Paillier scheme, the solution uses attribute based encryption for the publication payload. 
Key distribution uses a specific attribute-based group key management (AB-GKM) scheme~\cite{ABGKM}.
Subscribers can derive payload decryption keys using attribute credentials in their possession along with public key information.
This permits enforcing functional-encryption-based access control by allowing only restricted sets of subscribers to decrypt payloads.

\emph{Security considerations.} 
The blinding operations following the Paillier encryption are considered secure to chosen plaintext attacks (CPA).
The change of the private $\mu$ parameter in the Paillier cryptosystem into a public one is argued as safe in~\cite{Nabeel:2013}. 
The authors base their argument on the difficulty for an attacker to derive the other secret $\lambda$ parameter, needed along $\mu$.

\emph{Practical aspects.}
The drawback of the first two versions of the scheme~\cite{Nabeel:2012,Nabeel2009Privacy-Preserv} is the tight coupling between publishers and subscribers, which is unrealistic and impractical for several applications. 
The initial communication phase from the subscriber to the publishers must take place for \emph{every} subscription, which is no longer necessary with the context manager introduced in the latest iteration of the scheme~\cite{Nabeel:2013}. However, the solution still requires a loose level of coupling between publishers and subscribers, i.e., the security parameters needed for the Paillier encryption and blinding specific to each publication context (for a publisher and the subscribers interested specifically in the publications of that publisher). 

The matching complexity is as follows.
The broker must execute one $\text{\emph{diff}}$ comparison for each constraint, as displayed in Figure~\ref{Nabeel}.
This results in a matching complexity in O($|\mathcal{C}|$) operations per subscription.
While neither of the three papers presents a computational complexity analysis, the experimental evaluation in~\cite{Nabeel:2013} illustrates this linear scaling behavior in the number of constraints.

We consider that this solution presents interesting features in practice for a variety of use cases, including both our motivating e-Health example detailed in Section~\ref{sec:ehealth} as well as the stock market scenario referred in Section~\ref{sec:intro:case}.
However, it also has several drawbacks.
The scheme supports encrypted matching only for numerical values. 
This might be a problem for the e-Health use case, although for evaluating simple equality constraints, small domains like defined sets of patient names and injuries could be mapped to numerical values. 
The first two versions of the scheme explicitly support containment. 
Optimizations relying on this would be useful in scenarios where the system load is high like the stock market scenario.
On the other hand, the tight publisher/subscriber coupling in these first two versions impairs scalability. 
This is solved in the latest scheme iteration, but no longer addresses containment. The latest scheme version supports publication payload encryption using an ABE mechanism.
As in the case of~\cite{Ion:2012} this would be specifically useful in the e-Health scenario for securely transporting additional data like patient analysis results, X-rays, etc., to be decrypted only by selective categories of medical personnel.
This feature helps in a potential application of the scheme to the multi domain architecture in the e-Health use case.
Finally, the security analysis of the scheme gives stronger results than ASPE~\cite{Choi2010A-Privacy-Enhan} which makes it more reliable in case of highly sensitive applications like the stock exchange use case.

\medskip
\noindent
\textbf{\cite{Li2004An-efficient-sc}} presents a scheme that preserves the ability to determine shared prefixes between an encrypted publication attribute and an encrypted subscription constraint.
The data representation using bit prefixes presents similarities to the one used by~\cite{Ion:2012}.

\emph{Scheme mechanism.}
Subscription constraints are expressed through closed intervals over finite domains.
These intervals correspond to a set of admissible prefixes, and a publication attribute value that falls into the interval shares one of the prefixes.
Consider, as an example, the interval $i=[32,111]$.
Using one-byte representation, the set of prefixes corresponding to the interval is $I=\{\texttt{001*}, \texttt{010*},$ $\texttt{0110*}\}$.
Any value in the interval $c$ will have a common prefix with one of the elements of $I$.
For instance, value $a = 64 = {\texttt{01000000}}$ matches the prefix ${\texttt{010*}}$.

\begin{figure}
\centering
\includegraphics[scale=0.5]{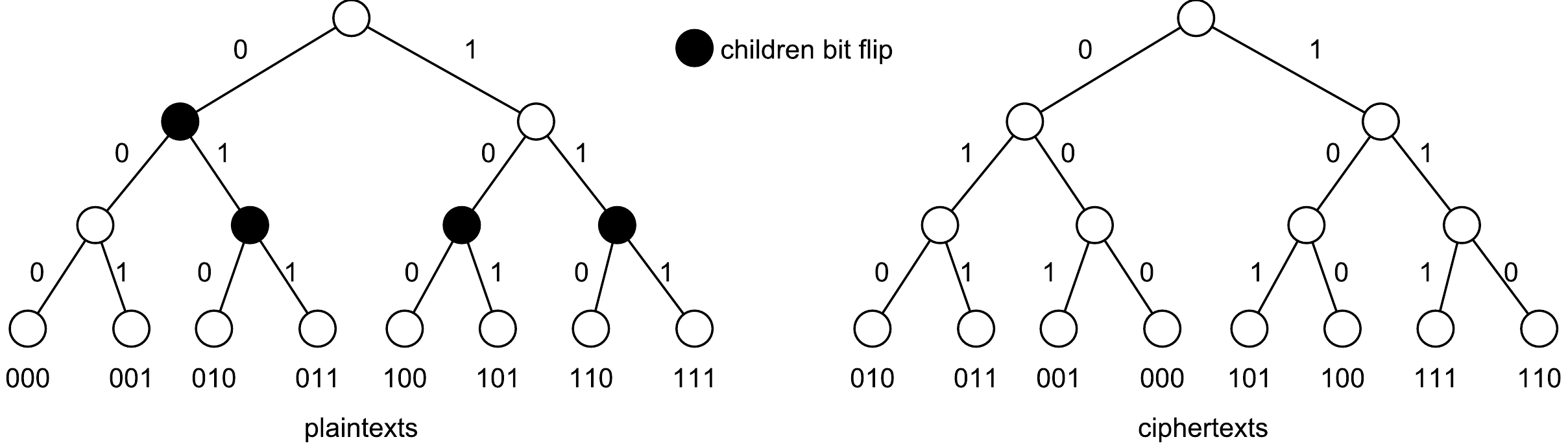}
\caption{Bit-flip based prefix-preserving permutation of~\protect\cite{Li2004An-efficient-sc}.}
\label{fig.lischeme}
\end{figure}

Attribute values and prefixes for the subscription constraints must be encrypted so that prefix matching remains possible.
This is achieved through a pseudorandom permutation of the bits.
Plaintext prefixes can be represented using a binary tree with a bit flip flag at each non-leaf node, as shown in Figure~\ref{fig.lischeme}.
When a flag is true (black nodes in Figure~\ref{fig.lischeme}), the bits of its children branches are flipped.
The complete tree permutation is the common key used by subscribers and publishers to encrypt constraint prefixes and attribute values. 
Matching between ciphertexts is thus the same operation as plaintext matching, and the technique is in essence a pre-mapped equality comparison.
The paper does not discuss containment determination, however it can be provided in a straightforward manner since the prefixes of a contained encrypted subscription will match the prefixes of its containers.

\emph{Security considerations.}
The authors mention that the scheme has limited resistance under a KPA attack model.
Subscription confidentiality and publication confidentiality can only be guaranteed under a weaker ciphertext-only-attack (COA). 

\emph{Practical aspects.}
The scheme requires a secret key to determine a common permutation for the publisher and the subscriber, but this key exchange is considered as an orthogonal problem. 

The matching complexity for each of the $c$ constraints is in $O(m)$, where $m$ is the number of bits used in the representation.
The total matching complexity is therefore in $O(|\mathcal{C}|\cdot m)$, which is less than the O($|\mathcal{C}|\cdot m^2$) complexity of~\cite{Ion:2012} that also relies on a bit token representation and prefix matching.
The difference is due to the maximum size of a constraint interval prefix set, which is $2(m-1)$ in~\cite{Li2004An-efficient-sc}.
Unfortunately, this performance advantage comes at the price of offering only COA security, which is a security limitation compared to the CPA-resilience guaranteed in~\cite{Ion:2012}.

\medskip
\noindent
\textbf{\cite{Tariq:2010,Tariq:2014}} consider confidentiality preservation in a peer-to-peer pub/sub architecture. 
The encrypted matching scheme is based on attribute-based encryption (ABE).
It supports numerical comparisons and prefix/suffix constraints on strings. 
The system model differs from a broker-based approach.
Each peer in the system participates to the collective operation of providing the pub/sub service to itself and other peers, as in~\cite{Choi2004HOMED:-A-Peer-to-Pee,Gupta2004Meghdoot:-content-ba,Voulgaris2006Sub-2-Sub:-Self}.
A peer can simultaneously act as a publisher and a subscriber.
Publications are delivered through an overlay connecting all subscribers and structured as a set of containment-based trees.
Each subscriber is responsible for matching incoming publications against its subscription, and to disseminate the publication according to containment rules.
Since the actual matching is performed by the subscribers themselves and not by an untrusted third party, the trust assumptions differ from a broker-based system as described below.

\begin{figure}
\centering
\includegraphics[scale=0.5]{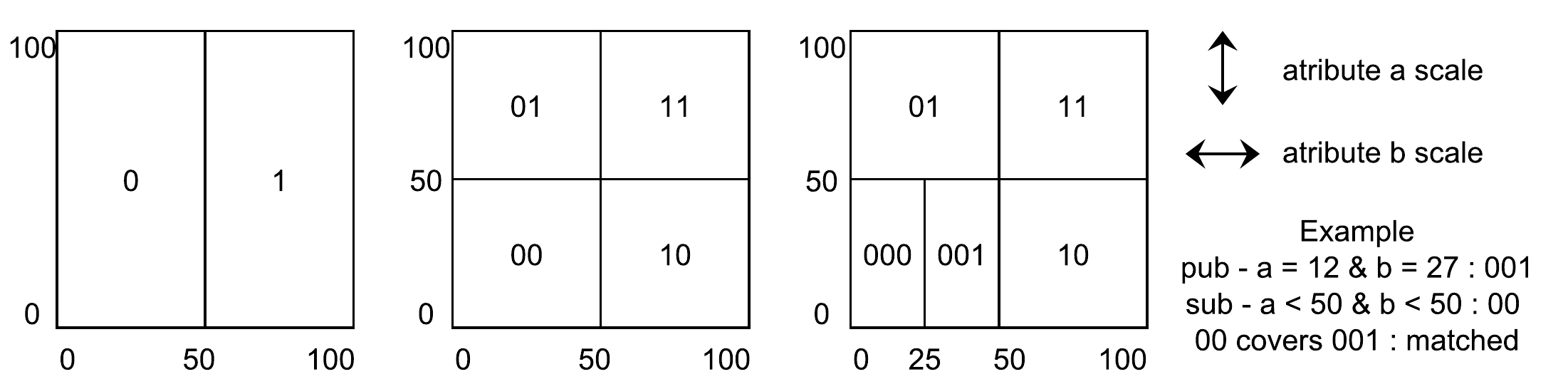}
\caption{Domain decomposition principle of~\protect\cite{Tariq:2014}.}
\label{fig.tariq}
\end{figure} 

\emph{Scheme mechanism.}
The encrypted matching scheme follows a pre-mapped equality comparison design.
Subscription constraints and publication attributes are mapped to bit strings.
For numerical values, the multi-dimensional attribute space is partitioned by splitting it alternatively over its dimensions.
Figure~\ref{fig.tariq} illustrates a domain partition for two dimensions.
Each slice maps to a bit string, reflecting which partition is considered at each step of the split.
While publications can be associated to a single slice, and thus to the longest possible bit string, subscriptions are associated to the smallest slice that contains their area of interest entirely.
Matching is performed by simple prefix matching: if publication $P=\{a=12,b=17\}$ corresponds to the bit string 001 and subscription $S=\{a<50 \wedge b < 50\}$ corresponds to bit string 00, $P$ matches $S$ as 00 is a prefix of 001.
Containment is determined similarly.
False positives are possible.
The matching and containment accuracy depend on several factors, such as the partitioning granularity, the generality or narrowness of subscriptions, and whether the range of the subscriptions overlaps the splitting points. As in~\cite{Raiciu2006Enabling-confid}, a good partitioning is fundamental to reach performance and scalability, and to avoid a high level of false positives.
String matching follows the same principle: suffix matching requires a mapping that considers strings in their reverse order.

\begin{figure}
\centering
\includegraphics[scale=0.53]{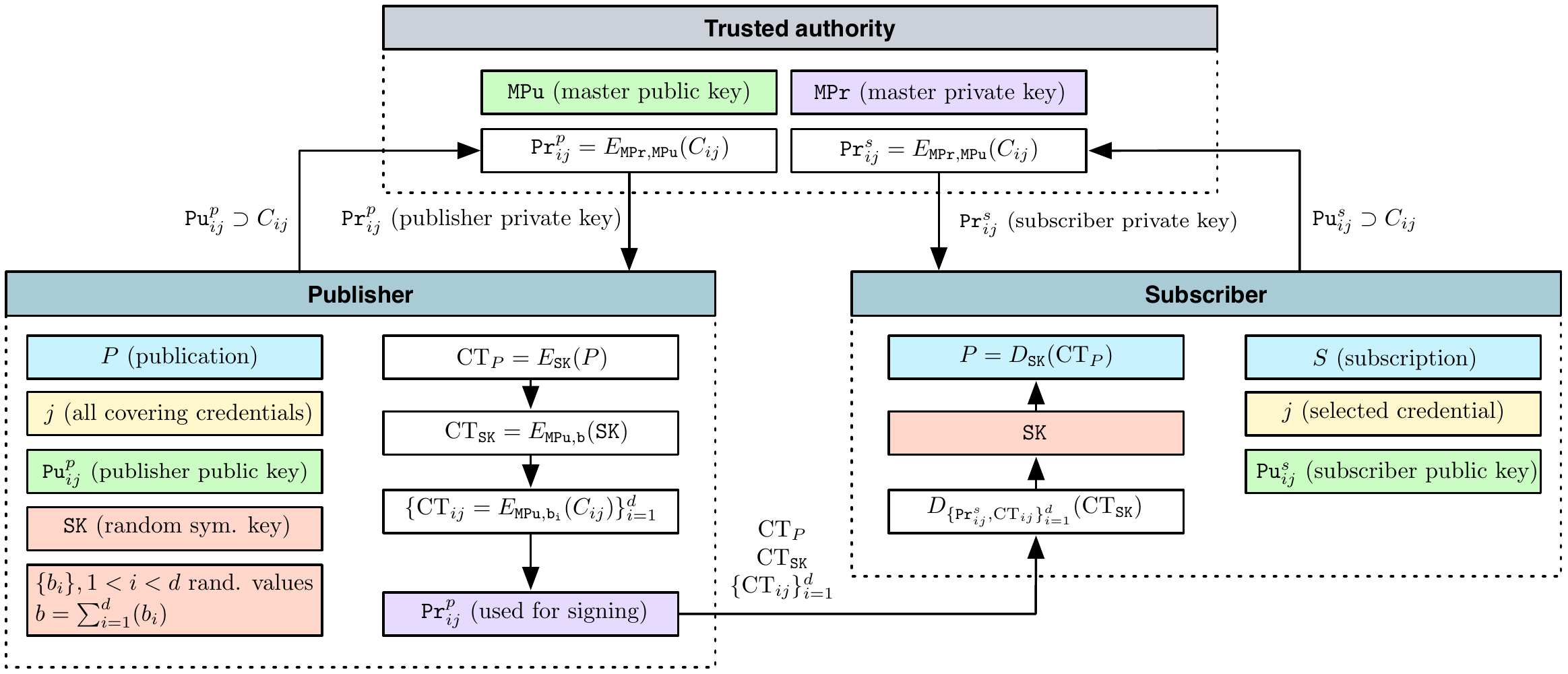}
\caption{Decryption scheme equivalent with matching functionality in~\protect\cite{Tariq:2014}.}
\label{fig.tariqdecrypt}
\end{figure}

Similarly as in~\cite{Ion:2012}, bit strings are used as \emph{credentials} for publications and subscriptions in an encryption scheme built upon ABE~\cite{Bethencourt:2007}.
We denote the credentials for publications and subscriptions as $C_{ij}$, where $i$ represents the attribute and $j$ the effective credential for that attribute (e.g., 001). 
We summarize the functional principles of the scheme in Figure~\ref{fig.tariqdecrypt}, and give details below.

The central and trusted authority acts as a key server.
It provides its master public key $\mathtt{MPu}$ for publication encryption.
The corresponding master private key $\mathtt{MPr}$ is used by the key server for generating specific private keys associated to credentials of publications and subscriptions as follows.

For each credential $C_{ij}$, a peer generates public keys $\mathtt{Pu}_{ij}^p$ and/or $\mathtt{Pu}_{ij}^s$.
These include the credentials for publication attributes and subscription constraints, respectively.
Peers then send these public keys to the key server.
Based on the credentials they contain, the key server generates private keys ($\mathtt{Pr}_{ij}^p$ and/or $\mathtt{Pr}_{ij}^s$) and returns them to the peer.

The private keys of a publisher are only used upon sending a publication for signature purposes.
A publisher first encrypts the publication $P$ with a locally generated secret symmetric encryption algorithm with key $\mathtt{SK}$ (e.g., using AES).
This yields the ciphertext $\text{CT}_{P}$. 
Then, it encrypts $\mathtt{SK}$ into $\text{CT}_{\mathtt{SK}}$ using $\mathtt{MPu}$ and some local random values $b_i$ generated for each attribute $i$ in the publication. 
These random values, along with parameters of $\mathtt{MPu}$, are used to generate another series of ciphertexts $\{\text{CT}_{ij}\}$, one for each credential and covering credentials $C_{ij}$ of each attribute in the publication.
Due to the inclusion of the covering credentials, any peer who subscribed to a matching criterion covering the publication will be able to match and decrypt it.

Ciphertexts $\text{CT}_{P}$, $\text{CT}_{\mathtt{SK}}$ and $\{\text{CT}_{ij}\}$ are sent to the pub/sub peer-to-peer overlay and disseminated towards subscribers following containment relations between subscribers.
Upon reception of the encrypted publication, subscribers first try to decrypt $\mathtt{SK}$ using the set of private keys $\mathtt{Pr}_{ij}^s$ that they obtained for subscription credentials and the set of $\{\text{CT}_{ij}\}$ they received along with the publication.
Then, subscribers can attempt to use $\mathtt{SK}$ to decrypt the actual publication.
This decryption can only succeed if the $\mathtt{SK}$ obtained is correct, which depends on the matching of the encrypted credentials.
This preliminary decryption of $\mathtt{SK}$ can be considered as the encrypted matching operation in the scheme. 
If the decrypted publication includes a specific padding or a hash that was appended before encryption, then the matching is positive, otherwise the obtained $SK$ is not correct and the matching is negative.

\emph{Practical aspects.}
Public key encryption is typically costlier than symmetric encryption; 
this justifies encrypting the actual publication content (which might be of arbitrary size) with a symmetric cipher and key $\mathtt{SK}$, and to rely on public key encryption only for the shorter plaintext $\mathtt{SK}$ itself.
However, decrypting all publications using $\mathtt{SK}$ is still required in order to effectively find the result of the matching operation. The decryption takes place even on peers that will eventually not be interested in the publication.

Decrypting $\mathtt{SK}$ requires the subscriber to use the private keys for each credential associated to the constraints in its subscription.
The article reports a decryption complexity in $O(d)$ per subscription where $d$ is the size of the attribute schema.
This is simply based on the number of multiplications done by the decryption algorithm, but does not include the cost of other component computations.

\begin{figure}[t]
\centering
\includegraphics[scale=0.5]{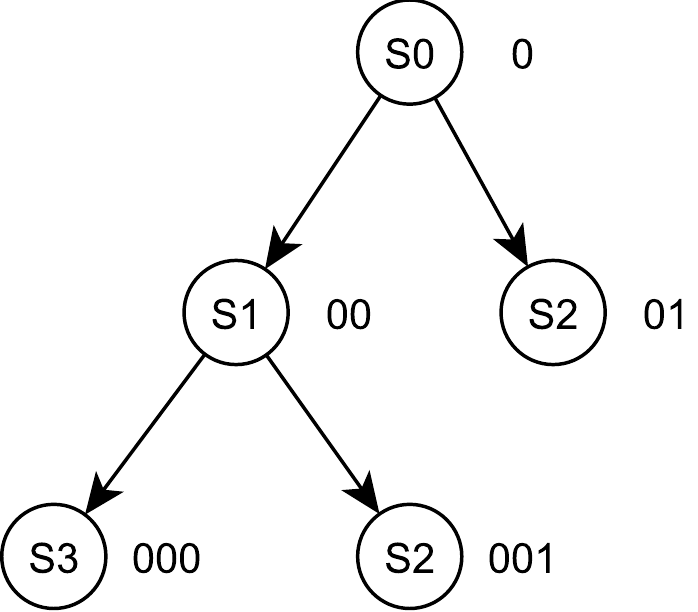}
\caption{Attribute tree in the~\protect\cite{Tariq:2014} scheme overlay.}
\label{fig.tariqtree}
\end{figure} 

The peer-to-peer system model where the service is provided through the interaction of ordinary peers rather than by a set of dedicated brokers leads to the following changes.
First, subscriptions are not effectively disseminated in the system but kept at the peer that emits them. 
Consequently, the cryptographic scheme does not include an actual subscription encryption and targets only a weak form of subscription confidentiality. 
Second, subscriber peers self-organize in an overlay structured as a set of trees.
Exploiting relations between subscriptions to obtain this self-organizing overlay structure is the main cause for the weak subscription confidentiality. 
Each tree in the overlay is based on containment relations between the subscriptions, with one tree per attribute in the schema. 
A peer thus belongs to multiple trees, and if it has several subscriptions with different constraints for the same attribute, it appears at different places in the same tree.
Figure~\ref{fig.tariqtree} presents an example attribute tree where a subscriber $S_2$ has two nodes with credentials 01 and 001 for the same attribute. 
The positions of the peers in the trees are established through connection request messages that are encrypted under the credentials of their subscriptions. 
Connection requests are gradually forwarded and decrypted by peers having matching subscriptions in the overlay until they successfully reach the closest possible nodes based on the attribute constraints.

We note that while the dissemination and matching of publications is decentralized, privacy preservation and encrypted matching rely on the presence of a centralized and omniscient trusted authority, departing from the peer-to-peer model.
Implementing such an authority as a decentralized system remains an open problem.

\emph{Security considerations.}
The encrypted matching operation does not need to guarantee publication confidentiality whenever matching is positive, since a peer matching a publication with one of its subscriptions is a valid destination of the subscription.
This is a fundamental difference with the broker-based model where the encrypted matching operation should not disclose the publication content, even when there is a successful match. 

The authors define a notion of weak subscription confidentiality and assume that leaking information about subscription containment is acceptable between peers that are linked in the overlay.
This is a natural assumption given the nature of the connection requests and the structure of the overlay. However, this containment support is not optional, which means that the scheme cannot be used for applications that prohibit containment information leaks. 
Relying on publication flooding to all peers in such cases would be intrinsically non-scalable, and does not seem to be a viable alternative.

\medskip
\noindent
\textbf{Other schemes.}
We conclude this section by discussing six contributions that do not have all the features of the schemes surveyed before in this section but nevertheless have characteristics or specificities worth mentioning.

\medskip
\noindent
\textbf{\cite{Srivatsa2005Securing-publis,Srivatsa2007Secure-Event-Di,Srivatsa:2011}} develop a pub/sub securing framework named \emph{EventGuard}. 
The framework includes a set of functionalities named \emph{guards}, which provide an extensive set of security properties.
Guards that address confidentiality concerns focus on protecting the payload, and not the headers of publications and subscriptions.
For this, EventGuard uses an architecture similar to the ABE technique in \cite{Tariq:2014} and \cite{Ion2010Supporting-Publ} that we covered earlier.
This architecture includes a key management solution that maintains a key tree per attribute.
The root of each tree is a key associated with the entire range of the attribute domain.
Each child node in the tree is a key for the corresponding partition of its parent's domain and can be derived from the parent key.
A subscriber can derive the key to decrypt a publication payload from a subscription key only if the publication matches the subscription.
The keys are disseminated in the system by a trusted centralized key distribution point.
The key management solution is mainly discussed in~\cite{Srivatsa2007Secure-Event-Di} for the case of numerical attributes, under the name of \emph{PSGuard} and extended in the later work.
String attributes and category hierarchies are covered in~\cite{Srivatsa:2011}.

For publication confidentiality and subscription confidentiality, the authors propose a tokenization technique similar to the one in~\cite{Song:2000} and previously discussed when we surveyed~\cite{Raiciu2006Enabling-confid}. This technique restricts encrypted matching to equality comparisons.

\medskip
\noindent
\textbf{\cite{EURECOM+2633}} presents a scheme based on multiple layer commutative encryption (MLCE).
Multiple layer encryption encrypts already encrypted messages one or more times. 
For two layers and two keys $k_1$ and $k_2$, the commutativity property guarantees that a plaintext $d$ is encrypted such that $E_{k_2}(E_{k_1}(d)) = E_{k_1}(E_{k_2}(d))$.

The scheme only supports equality comparisons, but on multiple attributes.
It is therefore classified in the more general class of content-based schemes, as topic-based schemes are restricted to those that allow filtering on a single attribute only.
Brokers are organized in a chain.
Each broker shares a different key with the other brokers in a set of $r$ predecessors and successors.
Upon receiving a message, a broker removes the cryptographic layers it can using its shared keys, performs the equality matching on what is left of the ciphertext, and adds a new encryption layer by using a key not shared by its $r-1$ following neighbors.
This allows the message to be transmitted further through the brokers while being protected at all times by at least one encryption layer that cannot be removed. 
While this design potentially fits any cryptographic algorithm with the appropriate commutativity properties, the authors focus on the Pohlig-Hellman scheme~\cite{Pohlig:1978}.

The interest of this article within the scope of this survey comes from the particular key distribution requirements.
Brokers, publishers and subscribers need to share secret keys with up to $r$ neighbors.
While the scheme allows choosing $r$ as small as 2, higher values of $r$ make the architecture more secure against a larger number of neighboring brokers colluding to obtain the plaintext.
This feature could be interesting for more general content-based pub/sub architectures.

\medskip
\noindent
\textbf{\cite{Shi:2007}}, while not specifically discussing confidentiality in pub/sub systems, presents another scheme similar to ABE-based solutions~\cite{Tariq:2014,Ion:2012}. 
The scheme targets multi-dimensional range queries over encrypted data and could be adapted to pub/sub architectures in order to selectively allow brokers to access subscriptions depending on the result of the matching operation.
The architecture allows the encryption of a query as an hyper-rectangle $B$, and allows its decryption only if a point of the data lies in $B$.
The use case considered is that of an investment broker assigned to execute a transaction order: buy or sell stock for an investor when an event matches the query $B$.
The investor does not trust the broker enough for revealing his query \emph{before} the order is executed. Therefore, only a positive match implies decrypting the query. 
As long as the matching fails (i.e., the order is not executed), the broker does not learn anything about the client's subscription.

\medskip
\noindent
\textbf{\cite{Pal:2012}} presents a pub/sub middleware solution where publishers and brokers rely on \emph{dissemination} and \emph{repository} servers for publication delivery.
Publishers encrypt the publications' payloads using ABE~\cite{Bethencourt:2007}, whereas the publications' headers are used to encrypt a unique \emph{id} for each publication through hidden vector encryption (HVE)~\cite{Boneh:2007}. 
The encrypted \emph{id}s are sent to the dissemination server, and the encrypted payloads are stored at the repository server.
The \emph{id}s are then forwarded to the subscribers, who perform the matching operation themselves.
Subscriptions are not disseminated and the scheme does not support their encryption.
When an encrypted publication header does not match any local subscription, a subscriber does not obtain any information.
When it does match, the subscriber obtains an identifier, which is then used to retrieve the payload from the repository server.
The payload is finally decrypted using ABE.

\medskip
\noindent
\textbf{\cite{Krishnan:2013}} presents an approach that relies on combining several building blocks in the areas of group algebra and boolean circuits. 
A first block consists in expressing subscription constraints through boolean circuits, which can be further transformed to a particular algebraic group program structure following a result obtained by \cite{Barrington:1986}. 
Publication headers are encoded in a similar group program structure. 
This representation for publications and subscriptions is further blinded using an approach proposed by \cite{Feige:1994} for secure computation of a public function over two private inputs.
The broker finally obtains the matching result using a group multiplication operation. 
Although seemingly promising in terms of expressivity, the article presents an initial work that is evaluated only by testing equality constraints and without a conclusive result on the variation of their number in a subscription. 
This is relevant considering that both the article and the work on which is based report severe drawbacks in the solution tractability depending on the function expressing the subscription, more precisely on the size of the boolean circuit.
The authors give several insights showing that the computational factor on the broker side can be reduced from exponential to polynomial, but without offering any detailed proofs.

\medskip
\noindent
\textbf{\cite{Crescenzo:2013}} presents an approach that is conceptually similar to the simple scheme used for equality filtering originally proposed by~\cite{Song:2000}. 
This simple scheme was used in the context of pub/sub in~\cite{Raiciu2006Enabling-confid} that we previously discussed. 
A third party introduced in the communication model replaces the untrusted broker, having the same role in the encrypted matching and being required to preserve the same privacy guarantees. 
The scheme relies on the use of two-layer cryptographic pseudonyms, which essentially are represented by a repeated application of a keyed pseudorandom function over the constraint values in a subscription and attribute values in a publication. 
The protocol is marginally different from the equality matching solution in~\cite{Raiciu2006Enabling-confid} including changes on the way the keys are used for the pseudorandom function, as well as details on the key exchange phase. 
\cite{Crescenzo:2013} provide more rigorous formal analysis that may be applicable for other schemes. 
However, the key exchange detailed implies a high degree of coupling and increases the communication overhead between the publisher and subscriber.
The two parties must indeed know each other in advance, and must exchange a new key for each subscription constraint, which inflicts a communication overhead.



\section{Challenges and unexplored issues}
\label{sec.conclu}

We presented a survey of techniques for providing confidentiality guarantees in publish/subscribe (pub/sub) systems.
We identified two main directions of research.
The first research direction focuses on the use and enforcement of security models, where confidentiality provisioning is facilitated by an access control solution (or some other specialized mechanism), controlling functions and rights for the system entities. 
In particular, this controls the access by brokers to the fields in subscriptions and publications used for matching and routing.
Solutions in this first direction require that the brokers accessing sensitive routable fields be trusted, forbidding their use in environments such as public clouds or shared infrastructures.
The second direction proposes domain-specific schemes performing \emph{encrypted matching}.
This allows an untrusted broker to determine whether an encrypted publication matches an encrypted subscription without the need to access the plaintexts.

In this section, we identify the main challenges and unexplored issues for confidentiality preserving pub/sub.
We believe that pub/sub systems will become an attractive option for a large number of real-life applications once these aspects are addressed satisfactorily.

\subsection{Key management}

Pub/sub systems, and in particular encrypted matching schemes, introduce two requirements for key management, and no available pub/sub system comes with a key management system that fulfills them.
Without proper key management, confidentiality cannot be provided in practice. 

The first requirement results from the decoupled nature of pub/sub communication.
Publishers do not know the destination of the publications, and subscribers do not know their origin.
As the destination of a message is unknown \emph{a priori}, prior key exchange between communicating entities is not feasible. 
We saw that solutions based on attribute-based encryption (ABE) such as~\cite{Ion:2012,Tariq:2010} could partially overcome this problem by associating keys with messages instead of associating them with system nodes.
However, even with this strategy, a trusted authority is required to provide some common information to the participants.
There exists research on decentralized ABE~\cite{decentralized_abe}, which does not require such a single trusted authority, but this was not yet applied in the area of pub/sub systems. 
In any case, the pub/sub solutions that rely on ABE, use it typically for publication payload protection, which leaves key exchange for message headers encryption as an open problem in respect to coupling.
One way to address this problem is to consider weaker decoupling assumptions, which is reasonable for many applications.
For instance, in the stock exchange scenario described in Section~\ref{sec:introduction}, one can imagine that a subscriber who registers a subscription for stock quotes is part of a larger agency that pays for the right to receive filtered publications.
In this setting, there is no association between the publisher and the particular subscriber host who will receive a publication.
Nonetheless, the stock market service is aware of the companies that paid for subscription rights as well as their subscriber domains.
In other words, the host-to-host pub/sub communication is still decoupled, but the relationships between the domains of the publisher and subscriber are known.
In this \emph{loosely coupled} setting, it is possible to adapt standard key management solutions for secure group communication.
An example of secure group management protocol is OFT (One-way Function Trees)~\cite{Sherman:2003}, suggested for pub/sub in~\cite{Bacon2008Access-control-}.
Secure group key management is a complex research area spanning a variety of techniques and protocol architectures~\cite{Rafaeli2003A-survey-of-key,Zou2005Secure}, and as such is beyond the scope of this survey.
This being said, we believe that adapting secure group communication for pub/sub systems is promising and should be studied further.

The second requirement is the need to refresh encryption keys. 
Key updates are needed when there are changes in the client trust, e.g., when a host using the current key is corrupted and must be evicted from the system.
Periodic key refreshing is also necessary to increase the resilience of the system to brute-force attacks.  
In content-based pub/sub, key updates introduce an important challenge: the invalidation of stored subscriptions when a new key replaces an old key.
All subscriptions stored by brokers in untrusted domains are encrypted with the old key.
As a result, they can no longer be matched against publications encrypted with the new key.
A naive solution would require that all subscribers re-encrypt their previous subscriptions with the new key, and resubmit each of them to the brokers.
This presents several major drawbacks.
First, it ties the completion of a key update phase to the network layer capabilities; 
when a large amount of subscriptions require resubmission, the key update phase can become prohibitively long.
The quality of service is also affected: handling subscriptions registration at the brokers results in higher load and network usage, which might ultimately impact the matching efficiency and increase the notification delays.
Another downside is that it forces subscribers to store their set of previous subscriptions.
Typically, a pub/sub service offers dependability guarantees (e.g., brokers storing subscriptions can have replicas, or even run more complex mechanisms handling recovery after failures).
If subscribers are also required to redundantly store subscriptions, they might need to develop or pay for another reliable storage service to handle failures. 
An appealing solution to this problem would be to develop extensions to existing encrypted matching schemes allowing secure re-encryption directly at the brokers, or to develop novel schemes supporting such a feature.
Such re-encryption solutions should however not impair the security of the scheme, i.e., the re-encryption token provided to brokers in untrusted domains shall not leak information about the original key, the new key, or the subscriptions themselves.
We note that the problem of re-encrypting elements stored in untrusted domains is also present in other contexts.
For encrypted databases~\cite{Popa:2012:CPQ:2330667.2330691}, encrypted data must be re-encrypted upon a key change, e.g., to prevent an evicted client from being able to further query the data.

\subsection{Confidentiality, performance and functional limitations}

Encrypted matching techniques impose a performance/confidentiality tradeoff.
In particular, the matching operation is significantly more costly and slower than plaintext matching.
As we have detailed previously, some encrypted matching schemes might also prevent from determining containment relations between subscriptions.
This restricts the use of efficient structures such as containment-based posets described in Section~\ref{sec:aspects:funcmodel}.
The lack of support for containment determination (or the possibility to decide whether to support it or not) can be a feature of the scheme, as the ability to determine containment may actually pose a threat to confidentiality by allowing to group similar subscriptions together.
This breaks indistinguishability and can allow determining the nature of a subscription based on statistical knowledge of its definition domain.
One possible way to overcome the resulting performance limitation is to rely on non-sensitive (and thus non-encrypted) fields in subscriptions to determine containment relations.
Another approach is to augment encrypted subscriptions and publications with compact structures allowing to \emph{pre-filter} subscriptions cheaply. The approach in~\cite{Barazzutti2012Thrifty-Privacy,Barazzutti:2015aa} proposes to augment subscriptions and publications with Bloom filters~\cite{Broder2002Network-Applica} encoding equality constraints for subscriptions and attributes values for publications. 
By allowing group membership comparisons, the filters allow knowing when a publication is sure not to match a given subscription.
This approach allows to discard a large fraction of subscriptions without using the costly encrypted matching function.
The Bloom filters raise the power of an attacker observing encrypted subscriptions in untrusted domains.
The studies in~\cite{SIROCCO2013,Barazzutti:2015aa} evaluate this power and propose solutions (e.g., Bloom filter truncation) to make it weak enough to be practically useful. Improving the performance of encrypted matching algorithms remains an important research problem.

Besides the unfavorable confidentiality/performance tradeoff of the existing encrypted matching schemes, there are also limitations in the expressiveness of the supported encrypted subscriptions, which handle mostly numerical values.
Strings constraints are only supported by a few schemes, and generally limited to prefix matching. 
Some schemes do not support range constraints for encrypted fields.
This limits the schemes to a hybrid expressive power, between that of topic-based and that of full-fledged content-based models.
We believe that a promising path to overcome these issues is to adapt other techniques developed for secure databases~\cite{Popa:2012:CPQ:2330667.2330691} or cryptography~\cite{Boneh:2007}.

\subsection{Attack model limitations and other security aspects}

Most of the techniques using encrypted matching consider confidentiality threats in a passive \emph{honest-but-curious} fashion.
In some practical scenarios, malicious entities might \emph{actively} attack the pub/sub system.
For instance, a malicious subscriber can try to corrupt several brokers running on some untrusted public cloud and increase the power of its attacks through the collusion of entities under its control. In our motivating example, collusion could also take the form of two hospitals trying to access data outside their respective domains. Although some of the schemes presented in this survey are resilient to some forms of active attacks and can somewhat mitigate collusion (e.g.,~\cite{Ion:2012,Choi2010A-Privacy-Enhan}), overall the resilience against active attacks is currently poorly understood and should be studied further.
The security models overviewed in Section~\ref{sec:models} are more flexible, since they are not typically constrained to a specific cryptographic algorithm and can use classic schemes proven secure for a wider range of scenarios -- but they obviously have the major drawback of not allowing filtering against sensitive fields in untrusted domains.

While confidentiality is the focus of this survey, we briefly describe other security aspects that apply to pub/sub systems.
They require attack models that generally complement the one considered in the literature.

First, \emph{integrity} relates to two distinct properties: \emph{message integrity} and \emph{origin integrity}~\cite{Bishop:2002}.
Message integrity is enforced when unauthorized modifications to the exchanged data can be detected.
Origin integrity (or authenticity) is the property of ensuring trustworthiness in the identity of the originator of a message.
When the attack model considers honest-but-curious brokers, both properties can be guaranteed for end-to-end messages using classical techniques such as computing and appending HMAC values to exchanged messages.
When the threat model considers stronger attackers such as malicious brokers, ensuring the integrity of subscriptions stored at the brokers, or the legitimacy of received publications, requires using specific techniques.
\cite{ChangMalicious} classifies the malicious threats faced by pub/sub systems.

Second, \emph{availability} is the property of ensuring continuous service or a certain quality of service to the clients under the presence of faults.
Such faults can either be crash faults or result from malicious actions like denial-of-service attacks (DoS).
For the former, \cite{Chang:2014:PFP:2611286.2611305} presents P2S, a replicated broker model using state machine replication based on the Paxos consensus protocol.
It considers a crash fault model for brokers with no malicious or byzantine behavior. 
For the latter, DoS attacks are a widely addressed security topic in distributed systems, but less often for pub/sub architectures.
\cite{Wun2007A-taxonomy-for-} gives a taxonomy of such attacks in pub/sub systems and potential directions for counteracting them.
Most often, DoS attacks are referenced as a side case or along other security properties ensured in more general pub/sub security models~\cite{Bacon2008Access-control-,Srivatsa2005Securing-publis}.

Finally, subscriptions \emph{anonymity} against colluding subscribers and brokers was studied  in~\cite{10.1109/TKDE.2012.177}. There is little work done on these important topics in the context of secure pub/sub, and we think that they should be studied further.

\subsection{Toward security models with encrypted matching}
\label{sec:modelsconclusion}

To conclude our presentation, we emphasize that given the current state of research, providing confidentiality in a pub/sub architecture without encrypted matching is much easier than by using encrypted matching, since well-established security models can be used and thoroughly tested algorithms can be deployed.
Furthermore, the proposed solutions in this direction generally come with a more holistic approach to security aspects such as key management, role management, etc.
However, without encrypted matching, confidentiality threats force us to significantly decrease the routing capabilities at the very core of content-based pub/sub systems, that is, in the set of brokers deployed on public and untrusted infrastructure.

Our primary conclusion is thus that research should advance towards integrating encrypted matching techniques into complete security solutions for pub/sub systems, taking into account its specific aspects like trust relations, key distribution and management, and performance impact.
We believe that only then will confidentiality-preserving pub/sub offer a significant advantage over more traditional communication systems based on coupled but secure communication.
As briefly discussed in Section~\ref{sec:confidentiality}, functional encryption might serve as basis for constructing a \emph{formal encrypted matching} framework.
Following the syntax from~\cite{functional_encryption_waters}, the matching in a pub/sub system could be formalized as $F:S\times P \to\{true, false\}$ where $S$ and $P$ are the sets 
of subscriptions and publications. 
An encrypted matching scheme would then be formally defined as a series of algorithms, such as key generation, subscription encryption, publication encryption, and matching that permits obtaining the functionality $F$ from the encrypted results.
Although this syntax can be easily derived from functional encryption, a rigorous and complete framework must take into account the particularities of encrypted matching, which have never been considered in the literature.
First, functional encryption is specifically defined for schemes based on public-key cryptography, which is only a subset of encrypted matching solutions.
Second, it does not consider pub/sub constraints like encrypting two types of messages (subscriptions and publications) resulting in different ciphertexts.
Third, the objective of predicate encryption, the main subclass of functional encryption, is to decrypt the ciphertext (matching is a precondition in this process).
This fails to capture the difference in trust between the entity that performs the matching (the broker) and the entity that decrypts (the subscriber).

Another important avenue of research is to bridge the gap between encrypted matching and other secure processing techniques like garbled circuits~\cite{DBLP:conf/focs/Yao86,Bellare:2012,DBLP:conf/eurocrypt/ZahurRE15} and homomorphic encryption~\cite{Gentry:2010:CAF:1666420.1666444}. 
The development of these techniques, like confidentiality-preserving pub/sub, is driven by the advent of cloud computing.
It however goes well beyond the operational need of matching between publications and subscriptions in a pub/sub service model, by aiming at performing \emph{general} processing on encrypted data, such as evaluating a function, without leaking sensitive information. 
Due to its more restricted functional nature, encrypted matching has much lower costs than even the most practical instances of homomorphic encryption, and it is an open problem to adapt the latter to the particularities of pub/sub (where only routing is required by the broker rather than decrypting, for instance).
We expect that encrypted matching schemes and homomorphic encryption algorithms will continue to co-exist and evolve independently in the near future.
However, considering the amount of resources currently devoted to homomorphic encryption and other related techniques, developers of encrypted matching schemes should pay close attention to new developments that might be applicable in pub/sub settings.
One should finally point out the recent developments around trusted execution environments like Intel SGX~\cite{sgx} that support secure execution in the cloud, and may also play a role in the development of efficient yet trustworthy confidentiality-preserving pub/sub systems.


\begin{acks}
\begin{wrapfigure}{r}{0.10\textwidth}
\includegraphics[width=\linewidth]{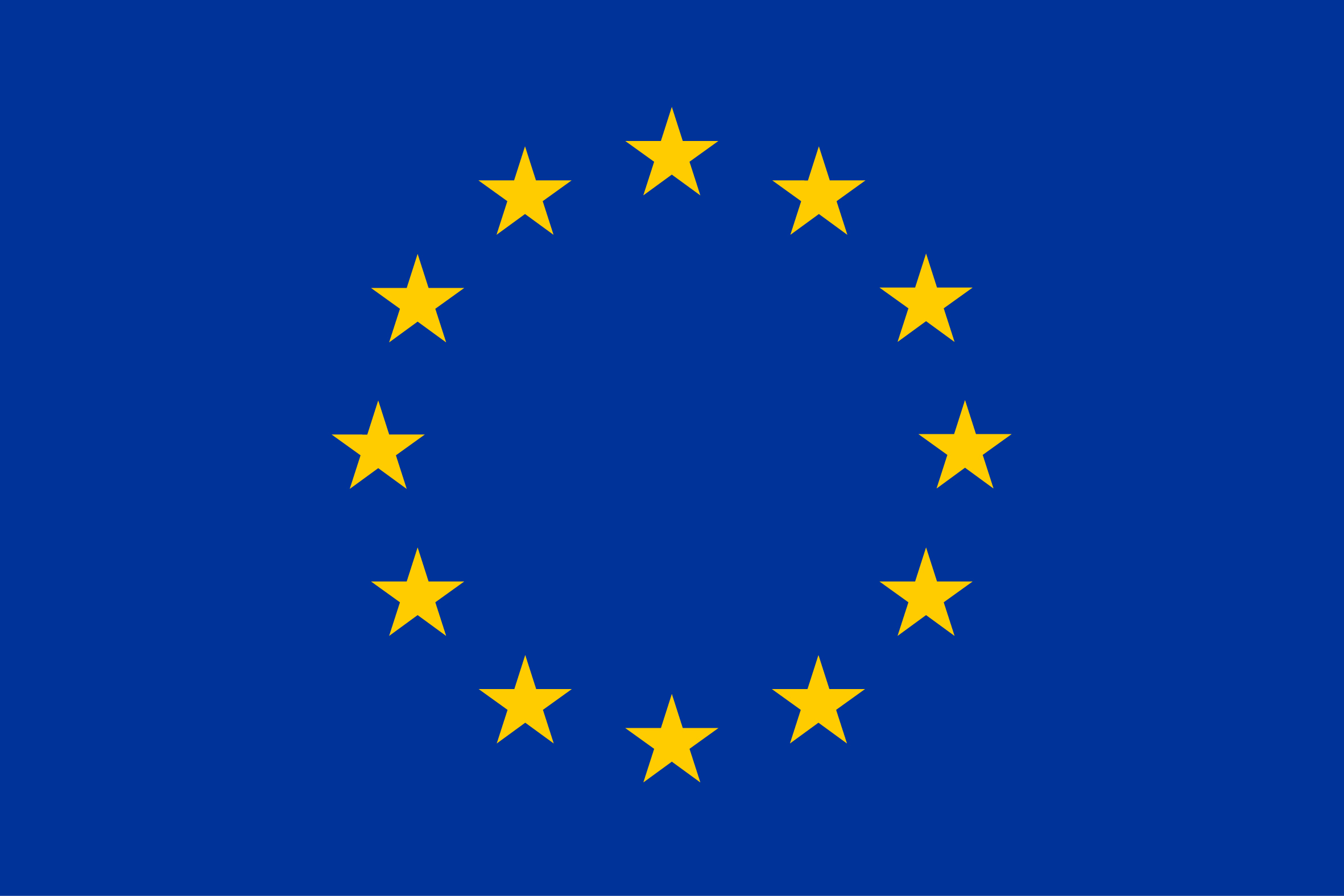} 
\end{wrapfigure}
The research leading to these results has partly received funding from the \emph{European Community's Seventh Framework Programme} (FP7/2007-2013) under grant agreement No 257843 (SRT-15 project). 
The dissemination of this work is partly funded from the \emph{European Union's Horizon 2020 research and innovation programme} under grant agreement No 692178 (EBSIS project).
\end{acks}

\footnotesize
\bibliographystyle{ACM-Reference-Format-Journals}
\bibliography{biblio}

\end{document}